\documentclass[sigconf, nonacm]{acmart}
\usepackage{graphicx}
\usepackage{caption}
\usepackage{float}  %
\usepackage{textcomp}
\usepackage{xcolor}
\usepackage{array}
\usepackage{ragged2e}
\usepackage{booktabs}
\usepackage{adjustbox}

\usepackage[ruled,vlined]{algorithm2e}
\usepackage{algpseudocode}
\usepackage{enumitem}
\usepackage{makecell}
\usepackage{multirow}
\usepackage{subfigure}
\usepackage{footnote}
\usepackage{threeparttable}
\usepackage{pifont}
\usepackage{colortbl}
\usepackage{flushend}
\usepackage{tcolorbox}
\usepackage{dsfont}
\tcbuselibrary{breakable,skins}

\usepackage{xcolor}

\colorlet{light-gray}{gray!45!white}
\colorlet{dark-green}{green!70!black}

\usepackage{marginnote}

\algnewcommand{\LineComment}[1]{\State \(\triangleright\) #1}

\usepackage{listings}

\definecolor{codegreen}{rgb}{0,0.6,0}
\definecolor{codegray}{rgb}{0.5,0.5,0.5}
\definecolor{backcolor}{RGB}{245,248,250}
\definecolor{emph}{RGB}{166,88,53}
\definecolor{nightblue}{RGB}{9,49,105}
\definecolor{keywords}{RGB}{207,33,46}
\definecolor{lightpurple}{RGB}{130,81,223}

\lstdefinestyle{mystyle}{
    backgroundcolor=\color{backcolor},
    commentstyle=\color{codegreen},
    keywordstyle=\color{keywords},
    stringstyle=\color{nightblue},
    basicstyle=\ttfamily\footnotesize,
    breakatwhitespace=false,         
    breaklines=true,                 
    captionpos=b,                    
    keepspaces=true,      
    numberstyle=\tiny\color{codegray},
    numbers=left,                    
    numbersep=2pt,                  
    showspaces=false,                
    showstringspaces=false,
    showtabs=false,                  
    tabsize=2,
    linewidth=1\columnwidth,
    frame=tb,
}
\newtheorem{definition}{Definition}  %

\newcommand{\set}[1]{\{#1\}}                    %
\newcommand{\len}[1]{|#1|}                      %
\newcommand{\indicator}{\mathds{1}}             %

\newcommand{\sql}[1]{\textup{\textsf{#1}}}
\newcommand{\ie}{\textit{i.e.},\xspace}

\newcommand{\ours}{\textsf{Operation-R1}\xspace}

\usepackage{balance}
\newcommand\vldbdoi{XX.XX/XXX.XX}
\newcommand\vldbpages{XXX-XXX}
\newcommand\vldbvolume{14}
\newcommand\vldbissue{1}
\newcommand\vldbyear{2020}
\newcommand\vldbauthors{\authors}
\newcommand\vldbtitle{\shorttitle} 
\newcommand\vldbavailabilityurl{https://github.com/ZJU-DAILY/Operation-R1.git}
\newcommand\vldbpagestyle{plain} 

\begin{document}
\title{Replacing Multi-Step Assembly of Data Preparation Pipelines with One-Step LLM Pipeline Generation for Table QA}

\author{Fengyu Li}
\affiliation{%
  \institution{Zhejiang University}
  \city{Hangzhou}
  \country{China}
}
\email{fengyuli@zju.edu.cn}

\author{Junhao Zhu}
\affiliation{%
  \institution{Zhejiang University}
  \city{Hangzhou}
  \country{China}
}
\email{zhujunhao@zju.edu.cn}

\author{Kaishi Song}
\affiliation{%
  \institution{Zhejiang University}
  \city{Hangzhou}
  \country{China}
}
\email{sks@zju.edu.cn}

\author{Lu Chen}
\affiliation{%
  \institution{Zhejiang University}
  \city{Hangzhou}
  \country{China}
}
\email{luchen@zju.edu.cn}

\author{Zhongming Yao}
\affiliation{%
  \institution{Northeastern University}
  \city{Shenyang}
  \country{China}
}
\email{yaozming@stumail.neu.edu.cn}

\author{Tianyi Li}
\affiliation{%
  \institution{Aalborg University}
  \city{Aalborg}
  \country{Denmark}
}
\email{tianyi@cs.aau.dk}

\author{Christian S. Jensen}
\affiliation{%
  \institution{Aalborg University}
  \city{Aalborg}
  \country{Denmark}
}
\email{csj@cs.aau.dk}

\begin{abstract}
Table Question Answering (TQA) aims to answer natural language questions over structured tables. Large Language Models (LLMs) enable promising solutions to this problem, with operator-centric solutions that generate table manipulation pipelines in a multi-step manner offering state-of-the-art performance. However, these solutions rely on multiple LLM calls, resulting in prohibitive latencies and computational costs.

We propose \ours, the first framework that trains lightweight LLMs (e.g., Qwen-4B/1.7B) via a novel variant of reinforcement learning with verifiable rewards to produce high-quality data-preparation pipelines for TQA in a single inference step. To train such an LLM, we first introduce a self-supervised rewarding mechanism to automatically obtain fine-grained pipeline-wise supervision signals for LLM training. We also propose variance-aware group resampling to mitigate training instability. To further enhance robustness of pipeline generation, we develop two complementary mechanisms: operation merge, which filters spurious operations through multi-candidate consensus, and adaptive rollback, which offers runtime protection against information loss in data transformation.
Experiments on two benchmark datasets show that, with the same LLM backbone, \ours achieves average absolute accuracy gains of 8.83 and 4.44 percentage points over multi-step preparation baselines, with 79\% table compression and a 2.2$\times$ reduction in monetary cost.
\end{abstract}

\maketitle

\pagestyle{\vldbpagestyle}
\begingroup\small\noindent\raggedright\textbf{PVLDB Reference Format:}\\
\vldbauthors. \vldbtitle. PVLDB, \vldbvolume(\vldbissue): \vldbpages, \vldbyear.\\
\href{https://doi.org/\vldbdoi}{doi:\vldbdoi}
\endgroup
\begingroup
\renewcommand\thefootnote{}\footnote{\noindent
This work is licensed under the Creative Commons BY-NC-ND 4.0 International License. Visit \url{https://creativecommons.org/licenses/by-nc-nd/4.0/} to view a copy of this license. For any use beyond those covered by this license, obtain permission by emailing \href{mailto:info@vldb.org}{info@vldb.org}. Copyright is held by the owner/author(s). Publication rights licensed to the VLDB Endowment. \\
\raggedright Proceedings of the VLDB Endowment, Vol. \vldbvolume, No. \vldbissue\ %
ISSN 2150-8097. \\
\href{https://doi.org/\vldbdoi}{doi:\vldbdoi} \\
}\addtocounter{footnote}{-1}\endgroup

\ifdefempty{\vldbavailabilityurl}{}{
\vspace{.3cm}
\begingroup\small\noindent\raggedright\textbf{PVLDB Artifact Availability:}\\
The source code, data, and/or other artifacts have been made available at \url{\vldbavailabilityurl}.
\endgroup
}

\section{Introduction}
Table Question Answering (TQA) aims to answer natural language (NL) queries over (semi-)structured tables, spanning spreadsheets, databases, web tables. It empowers non-technical users to explore and extract insights without writing formal query languages such as SQL, enhances analytical productivity, and enables applications in finance~\cite{zhu2025tableeval}, healthcare~\cite{bardhan2022drugehrqa}, and scientific research~\cite{DBLP:PramanickCV24}. For example, TQA for business intelligence enables users to query sales, customer, and financial tables using natural language (e.g., ``Which product had the highest growth in Q3?''), lowering the technical barrier to data analytics.

\begin{figure*}[t]  %
  \centering
  \includegraphics[width=\linewidth]{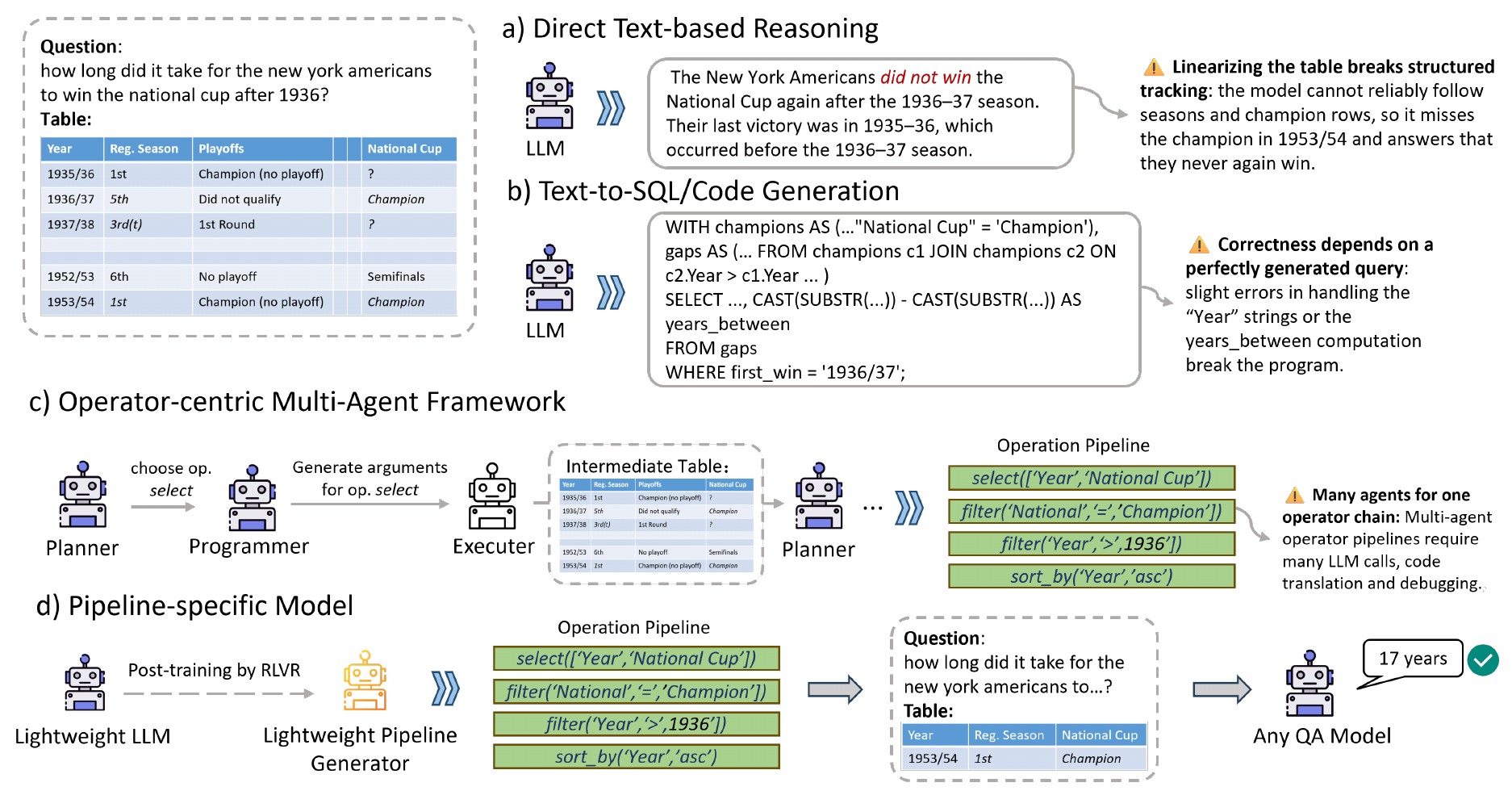}  %
  \caption{(a) Direct text-based reasoning, (b) Program-based reasoning, (c) Operation-based multi-agent framwork, and (d) Pipeline-specific model.}
  \label{fig:framework1}
\end{figure*}

Given their semantic understanding and instruction-following capabilities Large Language Models (LLMs) have advanced TQA substantially. As illustrated in Fig.\ref{fig:framework1}, one line of LLM-based TQA methods employ \emph{direct text-based reasoning} \cite{10.1145/3539618.3591708dater, wu2025tablebench}. These methods serialize tables into plain text and answer NL queries in a single step by reasoning directly over the text. However, they scale poorly to large tables because LLMs struggle to reliably identify relevant information within long contexts. Studies show that text-based reasoning deteriorates sharply as table size grows~\cite{fan2025autoprep}.

Another line of methods rely on \emph{Text-to-SQL/code generation}~\cite{10.1145/3539618.3591708dater,Cheng2023Binding,Zhang2024ReAcTable,rajkumar2022evaluating}. These methods translate NL queries into compositional executable programs (e.g., SQL or Python) and derive answers through program execution. Although often effective, they are fragile: LLMs easily introduce subtle errors into long, compound programs, causing executions to fail or produce corrupted results.

Motivated by limitations of above reasoning approaches, recent methods adopt \emph{operator-centric} approaches~\cite{wang2024chain-of-table-CoTable,fan2025autoprep}, where an LLM assembles a sequence of table manipulation operators (e.g., select, sort\_by) to restructure input tables into more compact, question-relevant tables that facilitate the TQA the most. To discover effective table manipulation operations, these methods employ multi-step reasoning, where table manipulation operations are generated step-by-step in the context of input questions, tables, and intermediate orchestration results. This operator-centric approach facilitates more transparent, interpretable reasoning and enables methods that achieve state-of-the-art performance.

Despite their accuracy and interpretability advantages, operator-centric pipelines incur substantially higher latency and computational cost due to repeated LLM invocations. This overhead, which is avoided by one-step approaches, is the primary barrier to practical deployment of these otherwise desirable operator-specific methods.

To enable the practical deployment of these methods, we introduce \ours, a framework that trains a lightweight LLM to generate a high-quality data preparation pipeline for TQA \emph{in a single inference step}. 
Inspired by the success of DeepSeek-R1~\cite{guo2025deepseek} and TableR1~\cite{yang2025tabler1inferencetimescalingtable}, we employ Reinforcement Learning with Verifiable Rewards (RLVR) to teach compact models (e.g., Qwen3-4B, Qwen-1.7B) to orchestrate well-structured data preparation tailored to each question–table pair.

However, building such a one-step orchestrator introduces several unique challenges not addressed by prior work. Below, we outline these challenges and how they motivate the key design components of \ours.

\textbf{\textit{Challenge I: Designing a data-preparation–aware reward mechanism for pipeline generation.}}
Training a lightweight LLM to generate effective data preparation pipelines requires reliable supervision signals that reflect pipeline quality. However, obtaining such signals is difficult because no ground-truth---or ``golden'' preprocessing pipelines'''---exists for TQA. The only available supervision in typical TQA datasets is final answers to each question, which is insufficient for directly supervising pipeline construction.

A natural surrogate is to evaluate a generated pipeline using an LLM-as-Judge~\cite{son2024llmasjudge}: execute the pipeline, present the transformed table to an LLM, and check whether it can infer the correct answer. Yet this approach has three critical drawbacks:
(i) Low accuracy. As shown in our experiments (Table~\ref{tab:main_results}, Section~\ref{sec:experiments}), even strong LLMs (e.g., Qwen3-8B) achieve under 80\% accuracy on WikiTQ when directly reasoning over tables, which means it may introduce noise into supervision signals.
(ii) Sparse feedback. The evaluation yields only a binary reward per pipeline, providing little insight into which operations are helpful or harmful.
(iii) High computational cost. Invoking an LLM-as-Judge for every pipeline drastically increases training cost.
Thus, a verifiable, fine-grained, and low-cost reward mechanism is essential.

To address this challenge, we introduce a self-supervised reward mechanism tailored to data preparation. We focus on \emph{Cell-Focused QA}, a setting where the answer appears verbatim as a table cell. This property allows us to verify each operation by checking whether it removes or preserves the answer-critical cell. Pipeline quality is then evaluated \emph{operation by operation}, enabling detailed attribution of failures and successes. To complement semantic correctness, we further incorporate an efficiency metric that measures how effectively the pipeline reduces table size. The combined reward provides fine-grained, stepwise feedback that directly reflects each operation’s utility, yielding informative and stable supervision for training the pipeline generator.

\textbf{\textit{Challenge II: Maintaining training stability under fine-grained reward supervision.}}
Although our approach introduces finer-grained supervision, integrating it with existing RLVR algorithms leads to training instability. In standard RLVR workflows, models are trained not with raw reward values but with \emph{normalized relative advantages}, which are computed by comparing a group of model responses for the same input. This normalization, while effective for coarse rewards, inevitably distorts the underlying learning signal. With our more granular pipeline rewards, this distortion becomes even more pronounced, ultimately misleading model optimization.
Our early analysis reveals two sources of distortion:
(a) Intra-group low-variance amplification. When reward variance within a sampled group is small, even noise-level differences are artificially amplified by the advantage normalization, causing the model to overfit to random fluctuations rather than meaningful reward differences.
(b) Inter-group reward distortion. Conversely, large and informative reward gaps across different groups are flattened after normalization, suppressing important learning signals that should guide the policy toward better pipelines.

To alleviate these issues, we propose a \emph{variance-aware group resampling} strategy that replaces the static grouping used in standard RLVR with a dynamic, variance-sensitive sampling process. The method imposes two constraints---each derived from the two distortion phenomena above---on sampled response groups. If any constraint is violated, the group is discarded and resampled. This ensures that each accepted group provides diverse, high-fidelity supervision signals, thereby preventing noise amplification and stabilizing model training.

\textbf{\textit{Challenge III: Ensuring robustness during preprocessing.}}
Even with a carefully finetuned LLM designed to generate high-quality pipelines, failures are inevitable when the model encounters unusual or previously unseen NL questions. A single faulty operation—such as overly aggressive filtering or an incorrect aggregation—can permanently remove question-critical information and make downstream reasoning impossible. This reveals a core tension between \emph{compression efficiency} and \emph{information preservation}: while aggressive table compression helps reduce reasoning complexity, any loss of essential data invalidates the entire preprocessing pipeline. The challenge, therefore, is to design a robustness mechanism that mitigates generation errors without compromising too much on efficiency and interpretability.

To address this, \ours incorporates two complementary mechanisms: \emph{Operation Merge} and \emph{Adaptive Rollback}.
Operation Merge tackles the nondeterminism inherent in pipeline generation by producing multiple candidate pipelines in parallel and constructing a final, more reliable pipeline through a consensus-based voting algorithm over an operation trie. This filters out spurious or unstable operations that may appear in single-pass outputs.
Adaptive Rollback provides runtime safeguards against information loss. When the downstream QA model signals that the processed table lacks sufficient information, the system incrementally reverts to earlier states---i.e., tables resulting from fewer preprocessing operations---ultimately reverting to the original table if necessary.

Together, these mechanisms strike a balance between the efficiency of single-pass preprocessing and the robustness required for real-world TQA. They enable our lightweight framework to reliably handle diverse questions and edge cases without relying on multi-agent coordination or iterative refinement.

The key contributions are summarized as below:

\begin{itemize}[topsep=0pt, leftmargin=*]
    \item To our knowledge, we present the first RLVR-based LLM post-training framework that enables lightweight LLMs to generate data preparation pipelines for TQA in a \emph{single inference step}.
    
    \item We introduce a self-supervised pipeline reward mechanism based on Cell-Focused QA that provides fine-grained, stepwise feedback for RLVR training. We further propose a variance-aware group resampling strategy to ensure stable optimization under such fine-grained rewards.
    
    \item We develop two complementary mechanisms to balance compression efficiency with information preservation: \emph{Operation Merge}, which filters spurious operations through multi-candidate consensus, and \emph{Adaptive Rollback}, which offers runtime protection against information loss.
    
    \item Extensive experiments on WikiTQ and TabFact show that \ours achieves absolute accuracy improvements of 9.62\% and 6.05\% over non-preprocessed baselines, respectively, while reducing table size by 79\% on average and requiring only a single lightweight model inference.

\end{itemize}

\section{PRELIMINARIES}

\subsection{Task Formulation}\label{sec:formulation}

\textbf{Tabular Question Answering (TQA).}
Let $q$ be an NL question and $T$ a table with columns (attributes) $\mathcal{C}={C_1,C_2,C_3,\ldots}$ and rows $\mathcal{R}={R_1,R_2,R_3,\ldots}$, where the cell at row $i$ and column $j$ is denoted by $c_{i,j}$. The goal of TQA is to generate a correct answer $a$ to question $q$ based on the content of table $T$.

Depending on the purpose of the question, TQA typically includes two problem types:
(i) \emph{Table-based Question Answering}: Given $q$ and $T$, the task is to generate a natural language answer $a$ or extract relevant cell values from $T$.
(ii) \emph{Table-based Fact Verification}: Given a statement $q$ and a table $T$, the task is to determine whether the statement is \emph{supported}, \emph{refuted}, or \emph{unverifiable} given the table.

\noindent\textbf{Question-Aware Data Preparation.}
Answering NL questions directly over an uncurated table is often ineffective and inefficient. The reasons are twofold:
(i) many table entries are irrelevant to the question, forcing the model to sift through large amounts of unnecessary information—the classic ``needle in a haystack'' problem; and
(ii) answers frequently require transformations or computations over the table rather than direct lookup, necessitating intermediate data preparation steps.

Thus, before generating answers, it is beneficial to apply a series of table transformation operations (data preparation operations, or simply \emph{operations}). Formally, let $\mathit{op}$ be an operation that transforms a table $T$ into a table $T'$, i.e., $T' = \mathit{op}(T)$. Given a pipeline $P = [\mathit{op}_1, \mathit{op}_2, \ldots, \mathit{op}_k]$, a curated table is produced as follows:
\begin{equation*}
    T'=\mathit{op}_k(\mathit{op}_{k-1}(\dots (\mathit{op}_2(\mathit{op}_1(T)))\dots))
\end{equation*}

We consider five data preparation operators. Based on prior studies~\cite{fan2025autoprep} and our analysis of benchmark and real-world tables, these operators cover most table manipulation needs in TQA. Below, we describe each operator and its parameters.

\textbf{\sql{Select.}}
The \texttt{select} operator removes columns from an input table.
Formally, given an input table $T$ and a set of column names $C$, $\text{select}_C(T) = T[C]$ denotes a processed table from $T$ by
retaining the only columns in $C$. This operator is useful when many columns are irrelevant to a question and may distract answering.

\textbf{\sql{Filter.}}
The \texttt{filter} operator selects rows that satisfy a given predicate.
Given an input table $T$, a column name $c$, a comparison operator $op \in \{==, >, <, \ldots\}$, and a threshold value $v$,
The result is denoted $\text{filter}_{c, op, v}(T)$.
For example, $\text{filter}_{\text{`Country'}, ==, \text{`USA'}}(T)$ retains rows where the country column cell equals `USA'.
This operation enables eliminating irrelevant tuples and retaining only those related to a question.

\textbf{\sql{Sort\_by.}}
The \texttt{sort\_by} operator sorts rows based on a specific column and optionally retains the top-$k$ rows.
Given an input table $T$, a column name $c$, an order $o \in \{\text{asc}, \text{desc}\}$, and an optional parameter $k$,
we denote the result by $\text{sort\_by}_{c, o, k}(T)$.
By default, when $k$ is omitted, all rows are returned, which we denote by $\text{sort\_by}_{c, o}(T)$.
This operation supports reasoning about ranking, extrema, and comparisons.

\textbf{\sql{Group\_by.}}
The \texttt{group\_by} operator groups rows by a specific column and counts the occurrences of each distinct value.
Given an input table $T$ and a column name $c$, we denote the result by $\text{group\_by}_c(T)$,
which produces a table with distinct values in column $c$ and their corresponding counts.
This operation is especially helpful for counting tasks that LLMs often struggle with.

\textbf{\sql{Add\_column.}}
The \texttt{add\_column} operator adds a new column inferred from existing columns according to a natural language description.
Given an input table $T$, a new column name $c'$ and a description $d$, we denote the result by $\text{add\_column}_{c', d}(T)$.
For example, $\text{add\_column}_{\text{`Gender'}, \text{`infer genders from the column Name'}}(T)$
adds a Gender column based on column Name.
This operator relies on semantic understanding and is implemented by invoking an LLM.

Operator \sql{add\_column} differs fundamentally from the other operators. We thus categorize the data preparation operators into two groups: structured operators, whose logic can be implemented by built-in or user-defined functions; and semantic operators that require reasoning over the data.

\subsection{RL with Verifiable Rewards}
\label{sec:prelim:rlvr}
\noindent\textbf{Reinforcement Learning with Verifiable Rewards (RLVR).} %
RLVR~\cite{lambert2024tulu3} is a family of LLM post-training methods in which an LLM is optimized through RL using automatically verifiable feedback as the reward signal. Unlike preference-model-based or human-rated rewards, RLVR rewards are computed by a verifier---an automated, deterministic, and programmatically checkable function that evaluates whether an output satisfies a well-defined success criterion. The verifier may check logical or syntactic constraints, unit-test outcomes, exact or semantic equivalence to reference solutions, or it may use other rule-based correctness conditions.
Concretely, given a prompt or input $x$ and a policy model (i.e., an LLM) $\pi_\theta$ producing an output $o$ (which may include a reasoning trace, or a chain-of-thought, and a final answer), a verifier computes a reward $r(o;x)$ (that is either a binary value $\{0,1\}$ or a continuous value in range $[0,1]$). %
RLVR uses an RL algorithm to update the policy model $\pi_{\theta}$ to maximize expected rewards. A typical optimization objective is formulated as follows:
\begin{equation}\label{eq:rlvr}
    \max_{\pi_{\theta}}
        \mathbb{E}_{x \sim \mathcal{D},o\sim\pi_{\theta}(\cdot\mid x)}\left[
            r(o;x) - 
            \beta\mathrm{KL}[
                \pi_{\theta}(\cdot\mid x)\Vert\pi_{\mathit{ref}}(\cdot\mid x)
            ]
        \right],
\end{equation}
where $\pi_{\mathit{ref}}$ is a reference policy (e.g. a pre-trained LLM) and $\beta>0$ controls the strength of the regularization. 
RLVR is applicable most straightforwardly in domains where outputs can be verified unambiguously, such as arithmetic reasoning, code synthesis, QA with known gold answers~\cite{guo2025deepseek,chen2025acereasonnemotronadvancingmathcode,he2025skyworkopenreasoner1}.

\noindent\textbf{Group Relative Policy Optimization (GRPO).} A member of the RLVR family, GRPO~\cite{guo2025deepseek,shao2024deepseekmath} has gained in prominence following the emergence of DeepSeek-R1. GRPO is a policy gradient algorithm designed to train an LLM in settings, where one can sample multiple candidate outputs per input and assign verifiable rewards to each candidate. Instead of relying on a learned value function or critic, GRPO uses within-group relative comparisons across the sampled outputs for the same input to compute advantages and update the policy. The use of the contrastive reward signal helps stabilize training when faced with sparse or binary rewards.

Formally, given an input $x$, GRPO samples a group $G$ of candidate outputs from the current (or ``old'') policy model $\pi_{\mathrm{old}}$, i.e., $G=\{o_i\}_{i=1}^{|G|}$. Each output $o_i$ is then scored by a verifier to produce a reward $R_i=r(o_i;x)$. Each output’s advantage is defined as the normalized difference from the group's mean reward:
\begin{equation}\label{eq:advantage}
    A_i=\frac{R_i-\mu_G}{\sigma_G},
\end{equation}
where $\mu_G=\frac{1}{|G|}\sum_{i=1}^{|G|}R_i$ is the group mean reward, and $\sigma_G=\sqrt{\frac{1}{|G|}\sum_{i=1}^{|G|}(R_i-\mu_G)^2}$ is the group standard deviation.

GRPO optimizes the policy model $\pi_{\theta}$ (i.e., an LLM) by maximizing the following token-level objective:

\begin{equation}\label{eq:grpo}
\begin{split}
    &\max_{\pi_{\theta}}\mathbb{E}_{x\sim\mathcal{D},\{o_i\}_{i=1}^{|G|}\sim\pi_{\theta}(\cdot\mid x)} \biggl[ 
        \frac{1}{|G|}\sum_G\frac{1}{|o_i|}\sum_{o_i}\Bigl\{ 
        \\
    &
        \min\bigl[\rho_{\mathit{i,t}}^{\theta} A_{\mathit{i,t}},
        \mathrm{clip}(\rho_{\mathit{i,t}}^{\theta},1-\epsilon,1+\epsilon)A_{\mathit{i,t}}\bigr]
        - \beta\mathrm{KL}(\pi_{\theta}\Vert\pi_{\mathit{ref}})
        \Bigr\}
    \biggr],
\end{split}
\end{equation}

\noindent where $\rho_{\mathit{i,t}}^{\theta} = \frac{\pi_{\theta}(o_{i,t} \mid x, o_{i,<t})}{\pi_{\theta_{\text{old}}}(o_{i,t} \mid x, o_{i,<t})}$ is the probability ratio at token position $t$, and $\epsilon$ is the clipping parameter that constrains policy updates. 
Although RLVR methods are prevalent for LLM post-training, they face problems, including a lack of verifiable rewards and unstable training, as discussed in the introduction, when they are applied to generate a high-quality data preparation pipeline, motivating the \ours approach.

\begin{figure*}[t]  %
  \centering
  \includegraphics[width=\linewidth]{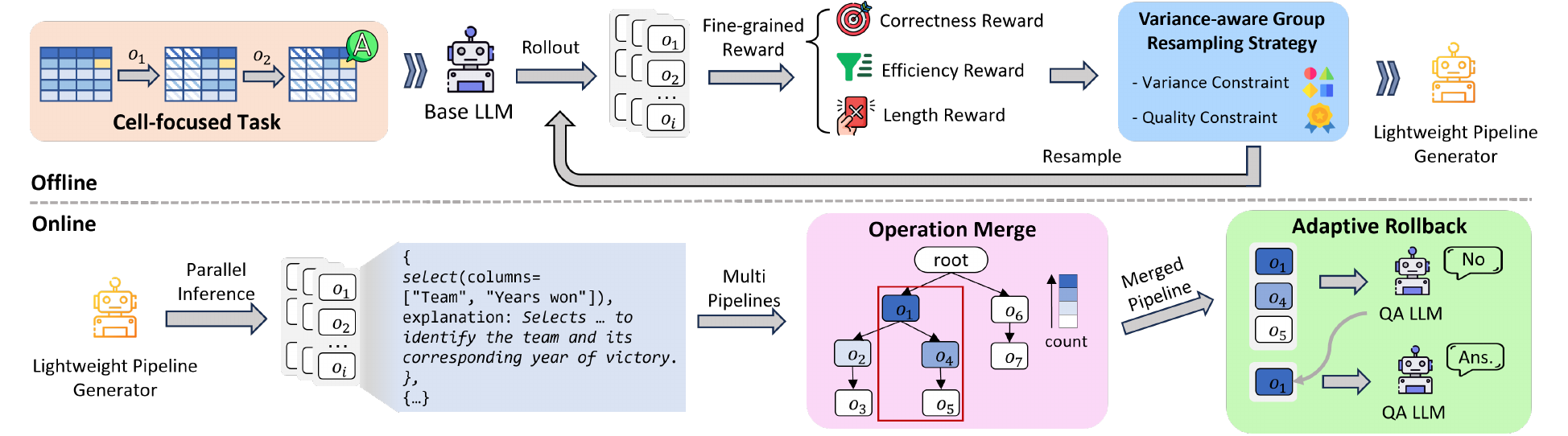}  %

  \caption{Overview of \ours, including ORPO-based offline training, parallel operator generation, operation-tree merging, and adaptive rollback for reliable tabular QA.}
  \label{fig:framework2}
\end{figure*}

\section{\ours}

\subsection{Framework Overview}
To bridge the gap between TQA and data preparation, we propose \ours, a question-oriented data-preparation orchestration framework for high-quality tabular question answering (Figure~\ref{fig:framework2}). Given a question and a table, \ours automatically composes and executes a tailored transformation sequence that reshapes the table to better expose the answer.
\textcolor{black}{
Unlike prior operator-centric methods that assemble pipelines through repeated LLM invocations, \ours generates the entire pipeline in a single forward pass, structurally eliminating multi-step orchestration overhead.}

As shown in Fig.~\ref{fig:framework2}, the offline stage of \ours aims to produce a TQA-oriented data preparation pipeline generator, built using open-sourced LLMs. To finetune an LLM for the task of QA-oriented data preparation, we propose \textsf{ORPO} (\underline{O}peration-wise Group \underline{R}elative \underline{P}olicy \underline{O}ptimization), a novel RVRL algorithm. \textsf{ORPO} features two designs that aim to make the post-training process more effective, efficient, and robust: (i) a self-supervised pipeline reward mechanism, providing finer-grained supervision signals tailored to TQA for model training; and (ii) a variance-aware group resampling strategy, enhancing the stability of the LLM training in case of reward collapse.

During the online stage, once receiving a natural language question $q$ over a table $T$, \ours leverages the data preparation pipeline generator derived from the offline stage to produce a sequence of operators. Each operator is organized as a JSON object containing the operator name, parameters, and an explanation (the usages of the operation). Due to hallucinations and nondeterminism in LLM inference, the \ours builds $N$ data preparation pipelines concurrently and then produces a consistent pipeline using an operation merge algorithm.

The JSON-represented operators are interpreted as physical operators and executed in the underlying environment (e.g., Python). Structured operators (e.g., \sql{select}, \sql{filter}) are compiled into code snippets and executed directly; semantic operators (e.g., \sql{add\_column}) are executed via LLM calls based on NL description.
This hybrid execution extends the capabilities of data preparation, expanding the scope of questions that TQA can address.

\textcolor{black}{Considering increasing cost of pipeline execution over massive tables, the system can be further accelerated with some minor modifications. For example, the operators can be executed on distributed engines, enabling parallel computing; the query optimization techniques (e.g., predicate pushdown) can be applied to reduce the unnecessary evaluations.}

Eventually, the processed table $T^{'}$ together with natural language question $q$ are be fed into a QA model to obtain the desired answer. To avoid the ``over-processing'' problem where key information is lost in the processing of data preparation, \ours also supports an adaptive rollback mechanism: When the QA model cannot identify relevant information from the table, the system will revert to the earlier table that contains more information.

In the following, we detail the post-training algorithm of \textsf{ORPO} and then the online question answering process.

\subsection{ORPO: Operation-wise Group Relative Policy Optimization}
As discussed earlier, there are numerous benefits to using a customized LLM to generate a QA-oriented data preparation pipeline, and GRPO performs well at customizing an LLM for a specific task~\cite{guo2025deepseek, yu2025dapoopensourcellmreinforcement}. 
The challenge however, is that as for data preparation operation generation requires assessing both final answers correctness and the quality of intermediate operations. We thus propose a low-cost surrogate operation reward mechanism to quantify the quality of generated operations in terms of both correctness and efficiency, forming the foundation for our \textsf{ORPO} training strategy.

\subsubsection{Self-supervised Pipeline Reward Mechanism} We first introduce the reward designs for operations in a generated data preparation pipeline in terms of correctness and efficiency.

\noindent\underline{\textit{Operation correctness rewarding.}} 
Ideally, a generated operation would be evaluated by comparing its output to a ground-truth preprocessed table; however, such a table cannot be obtained in advance; however, such a table cannot be obtained in advance.
We thus instead propose a surrogate operation rewarding mechanism that performs intermediate operation evaluations in a self-supervised manner on a ``cell-focused QA task''.

\color{black}
\begin{definition}[Cell-focused QA Task]
    Given a table $T$ containing a set of cell values $\set{c_\mathit{ij}}$, a question $q$ over $T$ is a cell-focused QA task if every value of the set $\mathcal{A}(q)$ of question answers comes from table $T$. Thus, the answer set $\mathcal{A}(q)$ formally satisfies $\forall a_k \in A(q) \; (\exists c_{i,j} \in T \; (a_k = c_{i,j}))$, where "=" denotes exact string match.
\end{definition}

\noindent\textit{Example.} The question in Figure~\ref{fig:framework1} (``How long did it take for the New York Americans to win the national cup after 1936?'') is \emph{non-cell-focused} as its answer ``17 years'' does not appear verbatim in the table. A cell-focused question over the same table would be ``What was the Playoffs result in the 1937/38 season?'', whose answer ``1st Round'' appears appears as a cell value.
\color{black}

On cell-focused QA tasks, we regard the operation evaluation as a judgment of whether the produced intermediate table contains the answers, which is a proxy for the comparison between the intermediate result and the virtual ground truth. 

Given a cell-focused QA task $q$ over an initial table $T$ and its answer set $\mathcal{A}(q)$, an LLM generates a sequence of operations $P=[\mathit{op}_1,\mathit{op}_2,\dots,\mathit{op}_n]$. The produced table of an intermediate operation $\mathit{op}_k\in P$ is denoted by $T_k$. The correctness of the operation $\mathit{op}_k$ is defined as follows.
\begin{equation}\label{eq:correctness}
    \mathit{cr}(\mathit{op_k},T_{k};\mathcal{A}(q))=\indicator(a=c,\forall a\in\mathcal{A}(q),\exists c\in T_k),
\end{equation}
where $\indicator$ is an indicator function. For simplicity, we omit $\mathcal{A}(q)$ from the notation of the operation correctness.

Based on the definition of correctness of a single operation, we propose a cumulative reward function to rate the correctness of a generated pipeline. Given a pipeline with $n$ data preparation operators, the accumulative accuracy reward assigned to the first $k\ (k\le n)$ steps in the pipeline is defined as follows.
\begin{equation}\label{eq:accuracy_reward}
    r_{\text{acc}}(k)=\frac{\sum_{i\le k}\mathit{cr}(\mathit{op}_i,T_{i})}{n}
\end{equation}

This accumulative accuracy reward function assigns higher scores to pipelines where every data preparation operation can retain answers to the question. While it encourages LLMs to generate  pipelines, lengthy pipelines are also encouraged, which is not what users want.

\noindent\underline{\textit{Operation efficiency rewarding.}} To avoid generating excessive data preparation pipelines, in addition to correctness, efficiency is considered as a reward to punish LLM generations that include redundant, useless operations in a pipeline.

Since one of the goals of data preparation is to eliminate irrelevant, noisy data from a given table, we define operator efficiency in terms of a compression ratio that measures reductions in the column and row sizes of the given table by an operation.

Without loss of generality, given an operator $\mathit{op}_k$ from a generated pipeline and its input table $T_{k-1}$ with columns $\mathcal{C}_{k-1}$ and rows $\mathcal{R}_{k-1}$, the table produced by $\mathit{op}_k$ is denoted by $T_{k}$ and has columns $\mathcal{C}_k$ and rows $\mathcal{R}_k$. The row-level and column-level compression ratio of operator $\mathit{op}_k$ is defined as $\frac{\len{\mathcal{R}_{k}}}{\len{\mathcal{R}_{k-1}}}$ and $\frac{\len{\mathcal{C}_{k}}}{\len{\mathcal{C}_{k-1}}}$, respectively.

Similar to the accumulative operator accuracy reward function, we define an accumulative operator efficiency reward function based on the compression ratio. As for an initial table $T$ with columns $\mathcal{C}$ and rows $\mathcal{R}$, the accumulative efficiency reward assigned to the first $k\ (k\le n)$ operations in the pipeline is defined:
\begin{equation}\label{eq:compress_reward}
    r_{\text{compress}}(k)= 0.5 \cdot \left(1 - \frac{|\mathcal{R}_k|}{|\mathcal{R}|}\right) + 0.5 \cdot \left(1 - \frac{|\mathcal{C}_k|}{|\mathcal{C}|}\right)
\end{equation}

\noindent\underline{\textit{Overall reward function.}} In addition to the operation correctness and operation efficiency rewards, we adopt a length reward from DAPO~\cite{yu2025dapoopensourcellmreinforcement} to constrain LLMs from generating overlong responses, alleviating ``over thinking'' ~\cite{sui2025stopoverthinkingsurveyefficient} as well as reducing latency and cost.

Formally, let $o$ denote an LLM response that contain a data preparation pipeline, and let $|o|$ be its token length. The length reward function on $o$ is defined as follows.
\begin{equation}
r_{\text{length}}(o) =
\begin{cases}
0 & |o| \le L_{\text{max}} - L_{\text{cache}} \\
\frac{(L_{\text{max}} - L_{\text{cache}}) - |o|}{L_{\text{cache}}} & L_{\text{max}} - L_{\text{cache}}<|o|\le L_{\text{max}} \\
-1 & |o| > L_{\text{max}},
\end{cases}
\label{eq:soft_punish}
\end{equation}
\textcolor{black}{where $L_{\text{max}}$ is set according to system constraints such as memory and rollout budget, and $L_{\text{cache}}$ is a slack token budget that defines the width of the soft tolerance region before strong penalties apply, thereby discouraging excessively long responses while avoiding abrupt penalties on near-limit samples.}

The overall reward function integrates the previous components with tunable weights:
\begin{equation}
    r(o,n)= r_{\text{acc}}(n) + \lambda_1\cdot r_{\text{compress}}(n) + \lambda_2\cdot r_{\text{length}}(o),
\end{equation}
where the weights ($\lambda_1$ and $\lambda_2$) balance operation correctness, operation and output efficiency, each of which are tuned according to performance on each datasets. Here we set $\lambda_1=\lambda_2=0.5$.

\subsubsection{LLM Post-training by \textsf{ORPO}} We integrate the self-supervised pipeline rewarding mechanism into the LLM post-training GRPO method to achieve the operation-wise group relative policy optimization method, \textsf{ORPO}. Given a question $q$ and a table $T$ from the training dataset, \textsf{ORPO} samples a group $G$ of LLM outputs from the current LLM $\pi_{\mathrm{old}}$, \ie $G=\{o_i\}_{i=1}^{|G|}$. Each output $o_i$ contains a data preparation pipeline $P_i=[\mathit{op}_1, \mathit{op}_2, \dots, \mathit{op}_{n_i}]$ and each output is then scored by the pipeline reward function $R_i=r(o_i,n_i)$. \textsf{ORPO} optimizes the LLM by maximizing the following token-level objective:
\begin{equation}\label{eq:orpo}
\begin{split}
    &\max_{\pi_{\theta}}\mathbb{E}_{(q,T)\sim\mathcal{D},\,\{o_i\}_{i=1}^{|G|}\sim\pi_{\theta}(\cdot\mid q,T)} \biggl[
        \frac{1}{|G|}\sum_G\frac{1}{|o_i|}\sum_{o_i}\Bigl\{
        \\
    &\min\bigl[
            \rho_{i,t}^{\theta} A^{\mathrm{ORPO}}_{i,t},
            \mathrm{clip}(\rho_{i,t}^{\theta},1-\epsilon,1+\epsilon) A^{\mathrm{ORPO}}_{i,t}
        \bigr]
        - \beta\,\mathrm{KL}(\pi_{\theta}\Vert\pi_{\mathrm{ref}})
        \Bigr\}
    \biggr],
\end{split}
\end{equation}

\noindent where $\rho_{\mathit{i,t}}^{\theta}=\frac{\pi_\theta(o_{i,t}|q,T,o_{i,<t})}{\pi_{\theta_{\text{old}}}(o_{i,t}|q,T,o_{i,<t})}$,
and $A^{\text{ORPO}}_{i,t}$ represents a normalized difference from the group mean operation reward, defined as:
\begin{equation}
A^{\text{GRPO}}_{i,t}
=
\frac{R_i - \mu_r}{\sigma_r + \epsilon},
\label{eq:orpo-adv}
\end{equation}
where $\mu_r$ and $\sigma_r$ are the mean and standard deviation of rewards among the sampled completions $o_i$ in the same batch, and $\epsilon$ is a small constant that ensures numerical stability.

\noindent\underline{\textit{Variance-aware group resampling strategy (VGR strategy).}}
Although \textsf{ORPO} introduces finer-grained supervision signals, it also causes more instability in the LLM training, which manifests itself in two forms. 
The first is \emph{low-variance amplification}: when rewards assigned to different LLM responses in a group are very similar (\ie the reward variance is close to zero), the group-wise normalization in \textsf{ORPO} amplifies tiny, often noise-level differences into large positive and negative advantages, causing the optimizer to overreact to spurious contrasts between almost indistinguishable trajectories and leading to unstable gradient updates.
Second, we observe a \emph{reward distortion problem}: in \textsf{ORPO}, the raw rewards $\{R_i\}$ are transformed into group-wise normalized advantages $A^{\text{ORPO}}_i$ (Eq.~\ref{eq:orpo-adv}), and the policy update is driven purely by such relative differences instead of by absolute reward values. This means that the model training does not receive real supervision signals.
For instance, two vastly different groups of absolute reward scores $[0.9, 0.5, 0.5]$ and $[0.3, 0.1, 0.1]$ will be normalized to the nearly identical advantages $[1.41, -0.70, -0.70]$ and $[1.43, -0.72, -0.72]$. Here training samples with a high reward of $0.9$ and with a low reward of $0.3$ across two groups will have the same effect (both are regarded as positive signals), consequently misleading the LLM training. This creates a critical vulnerability as highlighted in recent relative optimization studies~\cite{wang2025aipo,liu2026gdpogrouprewarddecouplednormalization}: the optimization process inadvertently reinforces the ``least bad'' sample in a sub-optimal group (e.g., the $0.3$ reward case) simply because it outperforms its peers locally, while failing to distinguish between relative superiority and absolute correctness.

\textcolor{black}{These two issues are largely absent in traditional GRPO applications such as mathematical reasoning or code synthesis, where binary correctness rewards naturally induce high intra-group variance and provide absolute oracle signals robust to cross-group normalization noise (e.g., ``least bad'')~\cite{mroueh2025reinforcementlearningverifiablerewards}. 
In our setting, data preparation considers both  compression efficiency and information preservation, inducing dense, continuous rewarding. Our task further demands precise operator composition---a single erroneous operator causes catastrophic downstream failure---and our lightweight models (1.7B/4B) trained on limited examples ($\sim$5K) could be more sensitive to noisy supervision than larger models, making robust supervision construction necessary.}

To address these challenges, we propose a \emph{variance-aware group resampling strategy}, which yields dynamic groups that provide diverse, informative training samples based on two group constraints.
Consider a group $G$ of LLM outputs containing various data preparation pipelines $G=\set{o_i}_{i=1}^{\len{G}}$. Each LLM output is assigned with a reward by the reward function, forming a group of rewards $\mathcal{R} = \{R_1, R_2, \dots, R_{\len{G}}\}$. The group mean reward $\mu_{G}$ is computed by $\mu_G=\frac{1}{\len{G}}\sum_{i=1}^{\len{G}} R_i$, and the reward variance $\sigma_G$ is calculated by $\sigma_G=\frac{1}{\len{G}}\sum_{i=1}^{\len{G}} (R_i - \mu_G)^2$. To overcome the two dilemmas discussed above, we have following two constraints to the group $G$:
\begin{itemize}[topsep=0pt, leftmargin=*]
    \item \emph{Variance constraint}: the reward variance of the group has to satisfy $\sigma_G\ge\epsilon$. This constraint avoids the \emph{signal collapse effect}.
    \item \emph{Quality constraint}: the maximum reward value $R_{\mathrm{max}}$ within the group has to satisfy $R_{\mathrm{max}}\ge\tau$, which guarantees that at least one output provides a high-quality data preparation pipeline. This constraint overcomes the \emph{reward distortion problem}.
\end{itemize}

\textcolor{black}{Once a group violates any of the constraints, it triggers group resampling \emph{within the same model update step}: the current group is discarded and a new group is generated for the same input before any gradient update is performed. This resampling repeats until a qualifying group is found, after which gradients are computed and the model is updated.}
Our variance-aware group resampling contributes to collecting a more diverse and higher-quality data preparation pipeline reasoning trajectories as the training data, and our empirical studies in Section~\ref{sec:ablation} validates that the resampling strategy enhances stability of LLM training under fine-grained reward regimes and mitigates early-stage reward collapse.

\subsection{Operation Merge}
In the online phase, the pipeline generator trained by \textsf{ORPO} is used to generate a sequence of data preparation operators tailored to the specific question and given table. To enhance the robustness and consistency of pipeline generation, inspired by~\cite{brown2024large}, \ours asks the generator producing $N$ pipeline candidates and then returns one of them with the most ``votes'' as the final data preparation pipeline. The vote for an operator represents how many pipeline candidates include that operator. The more votes a pipeline candidate has, the more reliable operators it contains. 

We propose an operation merge algorithm to count the votes and find the most trustworthy pipeline among all candidates efficiently.
The core data structure in operation merge is an operation trie built upon all pipeline candidates. In the operation trie, each branch represents a candidate pipeline, and each node in a branch represents an operator in the pipeline. If two candidate pipelines share the same operator sub-sequence, they share the same sub-branch in the operation trie. Each node in the trie has a weight recording how many candidate pipelines contain the operator (\ie the number of votes). With an operation trie, the problem of searching a pipeline that most LLM orchestrations advocate can be solved by identifying a branch of a trie with the maximum sum of node weights. In case of multiple branches having same node weights, a longer branch is preferred as the final pipeline.

The complete merging procedure is in Algorithm~\ref{alg:operation-merge}, which strikes a balance between inclusivity and precision: it captures all potentially informative schema modifications (via union and retention) while enforcing consensus for other transformations. Empirically, this approach has been shown to improve the consistency of downstream QA performance by mitigating the effects of noisy or unstable operator predictions from individual generations.

\begin{algorithm}[t]
\caption{Operation Merge}
\color{black}
\label{alg:operation-merge}
\DontPrintSemicolon
\SetKwInOut{KwIn}{Input}
\SetKwInOut{KwOut}{Output}

\KwIn{Candidate pipelines \(\mathcal{L} = \{L_1, L_2, \ldots, L_N\}\)}
\KwOut{The merged pipeline \(L^*\)}

\(L^* \gets \emptyset\)\;

\tcp*[l]{Step 1: Aggregate all \texttt{select} operations}
\(C_{\text{sel}} \gets \bigcup_{L_i \in \mathcal{L}} \{\text{columns in } \texttt{select}(\cdot) \text{ operators of } L_i\}\)\;
append \(\texttt{select}(C_{\text{sel}})\) to \(L^*\)\;

\tcp*[l]{Step 2: Build an operation trie over the remaining operators}
\(\mathcal{L}' \gets \{L_i \setminus \{\texttt{select()}\} \mid L_i \in \mathcal{L}\}\)\;
initialize an empty trie \(\mathcal{T}\) with root node \(r\)\;
initialize the node weight \(w(v) \gets 0\) for all nodes \(v \in V(\mathcal{T})\)\;

\ForEach{pipeline \(L_i' \in \mathcal{L}'\)}{
    insert \(L_i'\) into \(\mathcal{T}\) as a branch starting from \(r\)\;
    \ForEach{node \(v\) along the insertion path}{
        \(w(v) \gets w(v) + 1\)\;
    }
}

\tcp*[l]{Return a path with maximum node weights}
\(\mathcal{P} \gets \{P \mid P \text{ is a root-to-leaf path in } \mathcal{T}\}\)\;
\(P^* \gets \arg\max_{P \in \mathcal{P}} \bigl[\sum_{v \in P} w(v)\bigr]\)\;
append all operators from \(P^*\) to \(L^*\)\;

\Return \(L^*\)\;
\end{algorithm}

\subsection{Adaptive Rollback}
Although we design lots of strategies and techniques to avoid generating incorrect data preparation pipelines, it still remains risks of over-processing and information loss. I.e, essential information may be inadvertently removed from the table. To address this challenge, we design an adaptive rollback mechanism that enables self-correction when information loss is detected.

The rollback process is triggered when the QA model detects that the processed table $T^{'}$ loses the necessary information to answer the question and it outputs ``\texttt{No data available}''. This output serves as a self-diagnostic signal indicating the model's uncertainty under missing data conditions.

\textcolor{black}{
To balance effectiveness and efficiency, we propose binary rollback, which identifies the first state that supports correct reasoning through binary search. Given a table $T$ and a produced data preparation pipeline $P = [\mathit{op}_1, \mathit{op}_2, \dots, \mathit{op}_n]$, we define a series of rollback states (the produced tables in each states) as follows:
\begin{equation}\label{eq:rollback}
    T_{\mathrm{rb}}^{(k)} = \begin{cases}
        \mathit{op}_{l_k}\bigl(\cdots\mathit{op}_1(T)\cdots\bigr) & l_k \geq 1 \\
        T & l_k = 0
    \end{cases}, \quad l_k = \left\lfloor \frac{n}{2^k} \right\rfloor
\end{equation}
Initially, the full pipeline is applied and obtains an initial state. If the system emits ``\texttt{No data vailable}'', we recover the state at the middle position $\lfloor l_k / 2 \rfloor$ of the current pipeline, and check whether the LLM can reason out a correct answer based on the produced table. If not, the search continues for the first half sub-pipeline.
}

\section{EXPERIMENTS}\label{sec:experiments}

\begin{table*}[t!]
\renewcommand\tabcolsep{2.5pt}
\centering
\caption{Comparison of Table Understanding Methods under Unified Model Settings.
 ($^\dagger$Thinking mode is not evaluated for all baselines due to non-trivial overhead. More details and statistics are referred to extend version~\cite{li2026fullpaper}.)
}
\vspace{-1mm}
\label{tab:main_results}
\renewcommand{\arraystretch}{1.15}
\begin{adjustbox}{width=\textwidth}
\begin{tabular}{lcclcclcclcclcclcclcc} 
\toprule
\multirow{2}{*}{\textbf{Model}} & \multicolumn{2}{c}{\textbf{Llama3.1-8B}} & \multicolumn{1}{c}{} & \multicolumn{2}{c}{\textbf{Qwen2.5-7B}} & \multicolumn{1}{c}{} & \multicolumn{2}{c}{\textbf{Qwen3-4B}} & \multicolumn{1}{c}{} & \multicolumn{2}{c}{\textcolor{black}{\textbf{Qwen3-4B(Thk)$^\dagger$}}}  & \multicolumn{1}{c}{} & \multicolumn{2}{c}{\textbf{Qwen3-8B}} & \multicolumn{1}{c}{} & \multicolumn{2}{c}{\textbf{GPT-4.1 mini}}  & \multicolumn{1}{c}{} & \multicolumn{2}{c}{\textcolor{black}{\textbf{DeepSeek-V3.2}}} \\ 
\cline{2-3}\cline{5-6}\cline{8-9}\cline{11-12}\cline{14-15}\cline{17-18}\cline{20-21}
                                & \textbf{WTQ} & \textbf{TF}                &                      & \textbf{WTQ} & \textbf{TF}               &                      & \textbf{WTQ} & \textbf{TF}    &  & \textbf{WTQ} & \textbf{TF}    &                      & \textbf{WTQ} & \textbf{TF}  &  &     \textbf{WTQ} & \textbf{TF} & & \textbf{WTQ} & \textbf{TF} \\ 
\toprule
No-Prep                        & 49.10           & 77.08                           &                      & 50.69           & 71.99                          &                      & 46.20           & 74.95 & & \textcolor{black}{77.46} & \textcolor{black}{92.89}              &  & 46.87           & 76.58               &                      & 67.84           & 88.41  &  & \textcolor{black}{69.94} & \textcolor{black}{92.44}                  \\
Dater                           & 29.66           & 72.63                           &                      & 46.87           & 66.45                          &                      & 49.85           & 82.07     & & - & -            &  & 53.27           & 68.13               &                      & 53.91           & 64.52     &  &  \textcolor{black}{74.28} & \textcolor{black}{82.91}              \\
Binder                          & 50.44           & 74.87                           &                      & 36.34           & 77.83                          &                      & 50.14           & 79.13    & & - & -             &  & 55.31           & 81.93               &                      & 60.01           & 83.20   &  &   \textcolor{black}{54.26} & \textcolor{black}{86.19}               \\
ReAcTable                       & 42.94               & -                           &                      & 47.33                & -                               &                      & 53.38                & -      & & - & -                &  & 52.30                & -                    &                      &  60.15               &  -  &        & \textcolor{black}{70.95} & \textcolor{black}{-}                \\

CoTable                         & 40.56           & 75.59                           &                      & 46.20           & 80.43                          &                      & 57.50           & 81.72     & & \textcolor{black}{62.27} & \textcolor{black}{89.62}            &  & 52.21           & \underline{91.25}               &                      & 75.04           & 88.69  &  &  \textcolor{black}{72.12} & \textcolor{black}{90.96}                 \\
AutoPrep                        & 17.05           & 49.52                           &                      & 55.64           & 74.99                          &                      & 59.68           & 53.36   & & - & -              &  & 63.53           & 78.28               &                      & 72.81           & 88.24       &  &   \textcolor{black}{74.26}  & \textcolor{black}{85.95}           \\
Operation-R1-1.7B               & \underline{60.31}   & \underline{79.84}                   &                      & \underline{65.68}   & \underline{84.54}                  &                      & \underline{68.16}   & \underline{90.46}   & & \textcolor{black}{\underline{79.10}} & \textcolor{black}{\underline{93.03}}      &  & \underline{68.39}   & 90.42       &                      & \underline{76.06}   & \underline{92.19}    &  & \textcolor{black}{\underline{81.24}} & \textcolor{black}{\underline{93.33}}       \\
Operation-R1-4B                 & \textbf{64.27}  & \textbf{84.24}                  &                      & \textbf{69.66}  & \textbf{88.44}                 &                      & \textbf{70.72}  & \textbf{91.35}   & & \textcolor{black}{\textbf{81.63}} & \textcolor{black}{\textbf{93.47}}     &  & \textbf{71.06}  & \textbf{92.14}      &                      & \textbf{76.77}  & \textbf{93.87}     &  & \textcolor{black}{\textbf{83.75}} & \textcolor{black}{\textbf{94.17}}      \\
\bottomrule
\end{tabular}
\end{adjustbox}
\end{table*}

\subsection{Experiments Setup}\label{sec:setup}
\textbf{Datasets.} \textcolor{black}{We evaluate \ours using 4 datasets, WikiTQ~\cite{pasupat2015compositionalwikitq}, TabFact~\cite{chen2020tabfact}, FeTaQA~\cite{nan2022fetaqa}, and TableBench~\cite{wu2025tablebench}.}

\begin{itemize}[leftmargin=3.4mm]
\item\textbf{WikiTQ} is a dataset for table question answering, consisting of complex natural language questions over diverse Wikipedia tables, with 17{,}689 training and 4{,}344, test pairs covering wide variations in table structures and reasoning types.
\item\textbf{TabFact} is a table-based fact verification dataset built from Wikipedia, where each statement-table pair is labeled as \textit{True} or \textit{False}, emphasizing logical reasoning and consistency checking over tabular evidence.
\textcolor{black}{\item\textbf{FeTaQA} is a free-form table question answering dataset. Built from Wikipedia tables and derived from selected ToTTo examples, it emphasizes multi-cell evidence aggregation, reasoning, and long-form answer generation.}
\textcolor{black}{\item\textbf{TableBench} is a comprehensive TQA benchmark covering four reasoning categories, including Fact Checking (FC), Numerical Reasoning (NR), Trend Forecasting, and Chart Generation.}

\end{itemize}

\noindent\textbf{Baselines.} We evaluate \ours against 10 baselines.

\begin{itemize}[leftmargin=3.4mm]
\item\textbf{No-Prep} performs direct table QA without preprocessing, using the prompting template from Table-R1 (Yang et al.)~\cite{yang2025tabler1inferencetimescalingtable}.

\item\textbf{DATER}~\cite{10.1145/3539618.3591708dater} is an LLM‐driven decomposer that splits a large evidence table into salient sub-tables and factorizes complex questions into executable sub-questions via a parsing-execution-filling pipeline. 

\item\textbf{Binder}~\cite{Cheng2023Binding} is a program-centric approach that binds language models to symbolic execution (e.g., SQL/code), compiling model outputs into interpretable, executable programs.

\item\textbf{ReAcTable}~\cite{Zhang2024ReAcTable} is a ReAct-style TQA system that incrementally constructs intermediate table representations and invokes external SQL/Python executors for progressive data transformation and answer derivation.

\item\textbf{CoTable (Chain-of-Table)}~\cite{wang2024chain-of-table-CoTable} is a tabular reasoning system that treats an evolving table as the chain of thought, iteratively generating operations to update the table and expose verifiable intermediate results before producing a final answer.

\item\textbf{AutoPrep}~\cite{fan2025autoprep} is a question-aware data preparation multi-agent framework that decomposes complex table reasoning tasks into structured data preparation subtasks through collaborative agents, including a planner, a preprocessor, and an executor.

\item\textbf{Table-R1 (Yang et al.)}~\cite{yang2025tabler1inferencetimescalingtable} introduces two variants for end2end table reasoning.
\textit{Table-R1-SFT} is a supervised fine-tuning model trained on $\sim$48.5k carefully filtered reasoning traces generated by DeepSeek-R1 across three task types (TQA, TFV, FF-TQA), employing full-parameter fine-tuning.
\textit{Table-R1-Zero}, instead of being initialized from the SFT checkpoint, is trained directly from the same base model using $\sim$69k verifiable samples with the GRPO algorithm.
\textcolor{black}{To enable a controlled comparison under identical settings, we further introduce two additional variants, \textit{Table-R1-GRPO} and \textit{Table-R1-DAPO}, which apply GRPO and DAPO respectively to train an LLM to directly predict final answers. Both share the same backbone (Qwen3-4B, thinking mode), LoRA fine-tuning, and training data (WikiTQ) as \textsf{Operation-R1}.}

\item \textbf{Table-R1 (Wu et al.)}~\cite{wu2025tabler1teleai} proposes a region-based reinforcement learning framework built on the Table Instruct-RE dataset ($\sim$20k), where table regions are automatically generated by Deep-Seek-R1 to capture the minimal evidence areas necessary for solving each problem. It adopts two-stage training that first guides LLMs with SFT and then optimizes via Table-Aware Group Relative Policy Optimization (TARPO) with a mixed reward balancing region precision and correctness.
\end{itemize}

\noindent\textbf{Evaluation Metrics.} 
We use \textit{accuracy} to evaluate model performance on both datasets. 
For WikiTQ and TableBench, accuracy is computed as the percentage of questions where the predicted answer matches the ground truth. Following the evaluator from Binder~\cite{Cheng2023Binding}, we apply answer normalization to handle unit variations and list-based answers.
For TabFact, accuracy measures the percentage of statements correctly classified as True or False. For FeTaQA, we report BLEU score.%

\noindent\textbf{Implementation details.} Refer to Appendix~A in the extended version~\cite{li2026fullpaper}.

\vspace{-2mm}
\subsection{Overall Performance}

\begin{table*}[t]
\centering
\caption{Comparison of End2End Training-based QA Methods.
($\dagger$: results from the original studies due to unavailable model releases or GPU limits).}
\vspace{-1mm}
\renewcommand{\arraystretch}{1.15}
\label{tab:end2end}
\begin{tabular}{llcccc} 
\hline
\multicolumn{1}{c}{\textbf{Method}} & \multicolumn{1}{c}{\textbf{Base Model}} & \textbf{Train Data Size} & \textbf{Finetuning Method} & \textbf{WikiTQ} & \textbf{TabFact}  \\ 
\hline
TableGPT2$^{\dagger}$                            & Qwen2.5-72B                             & 2.36M    & Full-Parameter & 71.5            & 82.5              \\
Table-R1-SFT (Yang et al.)$^{\dagger}$           & Llama-3.1-8B-Instruct                   & 34K      & Full-Parameter & \textbf{83.8}   & 91.1              \\
Table-R1-SFT (Yang et al.)$^{\dagger}$           & Qwen2.5-7B-Instruct                     & 34K      & Full-Parameter & \underline{81.9}    & 89.9              \\
\hline
Table-R1-ZERO (Yang et al.)$^{\dagger}$          & Llama-3.1-8B-Instruct                   & 49K      & Full-Parameter & 81.2            & 87.6              \\
Table-R1-ZERO (Yang et al.)$^{\dagger}$          & Qwen2.5-7B-Instruct                     & 49K      & Full-Parameter & 79.8            & 87.6              \\
Table-R1 (Wu et al.)$^{\dagger}$                 & Qwen3-8B (Thinking mode)                                & 20K      & Full-Parameter & 73.87           & -                 \\
Table-R1-GRPO                                & Qwen3-4B (Thinking mode)                                & 11K      & LoRA           & 78.78           & 91.97             \\
Table-R1-DAPO                                & Qwen3-4B (Thinking mode)                                & 11K      & LoRA           & 79.08           & 92.24             \\
Operation-R1                        & Qwen3-4B (Thinking mode)                                & 5K       & LoRA           & \underline{81.63}   & \textbf{93.47}    \\
\hline
\end{tabular}
\vspace{-1mm}
\end{table*}

We conduct experiments on WikiTQ and TabFact across multiple backbones. As shown in Table~\ref{tab:main_results}, Operation-R1 consistently outperforms all compared baselines under each backbone setting.

\textcolor{black}{
\textsf{Operation-R1}-4B improves over baselines by $8.83\%$ on WikiTQ and $4.44\%$ on TabFact on average; the smaller \textsf{Operation-R1}-1.7B variant surpasses baselines by $6.12\%$ and $2.71\%$ respectively. 
Traditional program-based methods (Dater, Binder) have no clear advantage over No-Prep due to execution brittleness. Operation-based methods (CoTable, AutoPrep) exhibit strong backbone dependency: on smaller LLMs, they show no consistent advantage over No-Prep and occasionally underperform it---AutoPrep in particular suffers catastrophic degradation (e.g., Qwen3-4B/TabFact: $53.36\%$ vs.\ $74.95\%$ No-Prep) as smaller models generate buggy code that traps the system in futile self-debugging loops. 
With DeepSeek-V3.2 (>600B) as backbone, baselines benefit substantially from the stronger model---Dater and AutoPrep improve to around $74.3\%$ on WikiTQ---yet \textsf{Operation-R1}-4B still surpasses them by over 9 points ($83.75\%$), confirming that our training strategy contribute gains orthogonal to backbone capability: the operator orchestration policy is model-agnostic, and a stronger backbone directly translates into stronger \textsf{Operation-R1} performance.
Under Qwen3-4B thinking mode, No-Prep already reaches $77.46\%$/$92.89\%$ with test-time scaling, leaving limited headroom; \textsf{Operation-R1}-4B nonetheless delivers consistent improvements ($+4.17\%$/$+0.58\%$). 
Thinking mode is not evaluated for multi-step baselines as adapting their prompting pipelines requires non-trivial redesign and incurs $\times6.8$ average overhead (more statistics are provided in Appendix B.5~\cite{li2026fullpaper}).
}

\textcolor{black}{
Operation-R1's stable performance stems from framing table preprocessing as a trainable operator orchestration task. By learning to compose operator sequences via ORPO, the system achieves robust results across backbones without the brittleness of code execution or the overhead of iterative agent collaboration.
}

Table~\ref{tab:end2end} compares Operation-R1 against state-of-the-art end-to-end training approaches. Despite using only 4B parameters and 5K training samples, Operation-R1 achieves competitive WikiTQ accuracy (81.63\%) and state-of-the-art TabFact performance (93.47\%).

\noindent\underline{\textit{Operator Generation vs. Natural Language Reasoning.}} Against large-scale supervised methods, Operation-R1 outperforms TableGPT2 (72B, 2.36M samples) by 10.12 points on WikiTQ and 10.97 points on TabFact, demonstrating the superiority of targeted operator-based preprocessing over massive generic pre-training. Table-R1-SFT with Llama-3.1-8B achieves the highest WikiTQ score (83.8\%) but requires 34K curated reasoning traces from DeepSeek-R1. Operation-R1 maintains robust performance with 6.8× less training data while surpassing all SFT variants on TabFact (+2.37 points). 
This provide empirical evidence that structured operator composition can be easier to learn than free-form textual reasoning in our setting.

\noindent\underline{\textit{Preprocessing-then-QA vs. End-to-End RL.}} 
Our controlled comparison with Table-R1-GRPO and Table-R1-DAPO—which train end-to-end QA on the same Qwen3-4B backbone with 11K samples—yields 78.78\%/91.97\% and 79.08\%/92.24\%. \textsf{Operation-R1} outperforms both by over 2 points on WikiTQ using only 5K samples, \textit{despite being trained solely on operator generation rather than the QA task itself}. This counterintuitive result stems from a fundamental difference in supervision quality: end-to-end QA conflates table understanding, information selection, and multi-step reasoning into a single objective with sparse, binary reward signals. Our two-stage paradigm decouples these challenges—Cell-Focused QA verification provides deterministic, step-wise feedback at each transformation, enabling precise credit assignment that makes operator learning tractable for lightweight models without the supervision complexity of monolithic approaches.

\vspace{-3mm}
\subsection{Efficiency Analysis}

To ensure fair comparison, we evaluate efficiency on the full test sets under controlled settings. 
\textcolor{black}{We measure end-to-end latency and actual API cost rather than token count, as API cost reflects LLM-side behavior such as reasoning effort and cache utilization.}
For latency, due to API stability concerns, we uniformly deploy Qwen3-8B across all baselines and Qwen3-4B (Thinking mode) for No-op and \textsf{Operation-R1}; for cost, we report actual API spending incurred by GPT-4.1-mini, while \textsf{Operation-R1} is costed using the Qwen3 API pricing on the DashScope platform.

\begin{figure}[t]
  \centering
  \includegraphics[width=\linewidth]{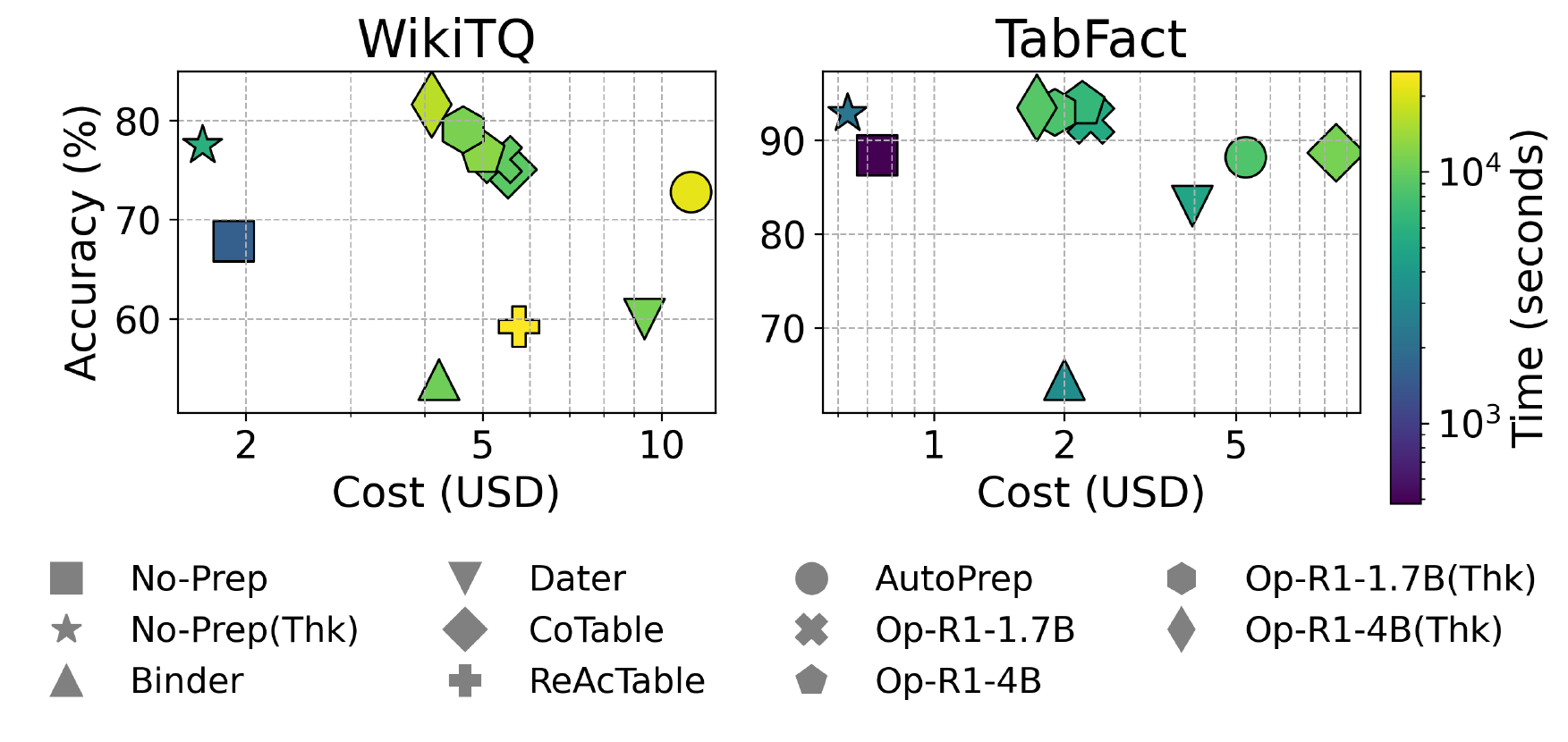}
  \vspace{-8mm}
  \caption{Efficiency-Accuracy-Cost Trade-off Comparison.}
  \label{fig:time_cost}
  \vspace{-5mm}
\end{figure}

\color{black}
Fig.~\ref{fig:time_cost} presents a comprehensive efficiency-accuracy-cost analysis across WikiTQ and TabFact. \textsf{Operation-R1} consistently achieves a favorable trade-off: in non-thinking mode, both the 1.7B and 4B variants outperform all multi-step agentic baselines in accuracy while incurring substantially lower latency and cost. Among the baselines, multi-step agentic baselines (AutoPrep, ReAcTable) impose the heaviest overhead due to their iterative LLM invocation loops, yet still fall short of \textsf{Operation-R1}'s accuracy. Program-based methods (Binder, Dater) scatter across the low-accuracy region without offering meaningful efficiency advantages. 

Enabling thinking mode roughly doubles latency but delivers consistent accuracy gains, representing a worthwhile test-time scaling trade-off that remains more cost-efficient than multi-step agentic alternatives at comparable accuracy levels. It is also worth noting that since Qwen3-1.7B and Qwen3-4B share identical per-token pricing on DashScope, local deployment of the 1.7B model would yield even lower inference cost in practice.
\color{black}

\subsection{Ablation Studies}
\label{sec:ablation}

\begin{figure*}[t]  %
  \centering
  \includegraphics[width=\linewidth]{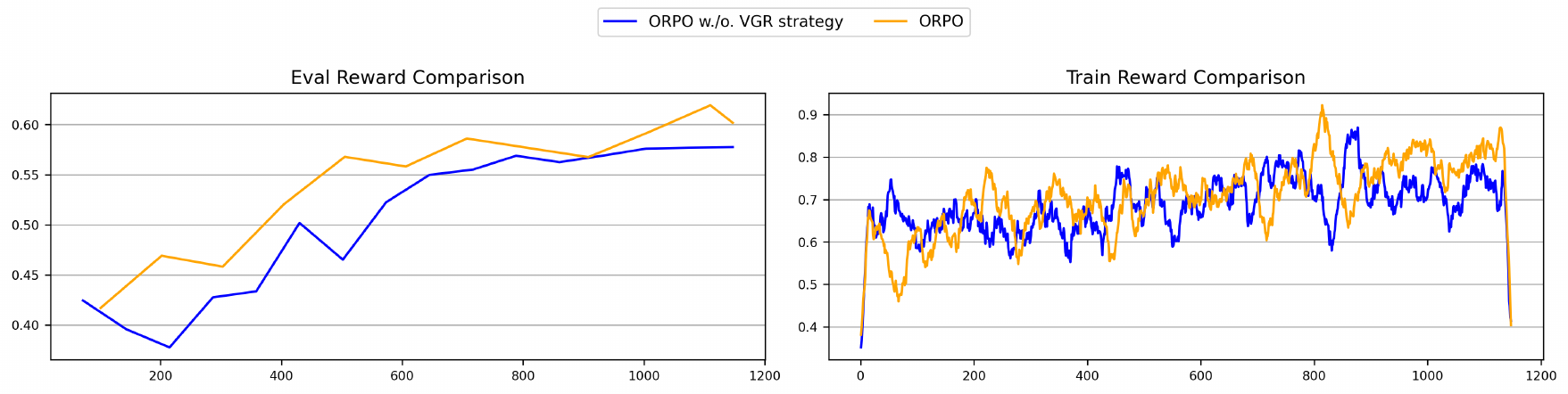}
  \vspace{-5mm}
  \caption{Training curves of Op-R1-1.7B with and without VGR strategy. Removing VGR causes a pronounced early drop in reward and slower subsequent improvement, whereas the VGR yields smoother, steadily increasing curves and higher rewards.}
  \label{fig:aggressive}
  \vspace{-2mm}
\end{figure*}

\textcolor{black}{
To validate each core component of Operation-R1, we conduct ablation studies on \ours with two backbone models. The results are presented in Table~\ref{tab:ablation}.
}

\textcolor{black}{
Completely removing ORPO training leads to an average accuracy drop of $6.8\%$ accuracy drops across both backbones: $-5.78$/$-2.67$ on Qwen2.5-7B and $-7.76$/$-3.94$ on Qwen3-4B (WikiTQ/TabFact), confirming that ORPO reduces operator generation errors at the modeling level. 
Disabling Operation Merge results in substantial performance drops of $4.2\%$ on the two backbones respectively ($-1.11$/$-2.67$ and $-4.79$/$-4.98$), confirming that consensus-based operator aggregation is critical for filtering spurious single-inference outputs. 
Similarly, removing Adaptive Rollback decreases accuracy by $5.8\%$ on average, demonstrating its role in handling over-processing failures at runtime. 
As observed, Operation Merge and Adaptive Rollback play critical roles in providing robust and high-quality data prep pipeline, serving as a robust system foundation atop a well-trained model rather than patches for recurring generation errors—a role complementary to, not in competition with, the modeling-level correction provided by ORPO.
}

\textcolor{black}{
When only VGR is disabled, accuracy drops by $0.60$/$1.99$ on Qwen2.5-7B and $1.06$/$0.19$ on Qwen3-4B, indicating its importance in maintaining training stability.} As shown in Fig.~\ref{fig:aggressive}, removing VGR leads to a pronounced early decline in reward curves and slower subsequent recovery, whereas the full VGR strategy yields smoother, steadily increasing curves and higher final rewards.

\begin{table}
\centering
\setlength{\abovecaptionskip}{1mm}
\caption{Ablation Studies on Two Backbone Models}
\label{tab:ablation}
\renewcommand{\arraystretch}{1.15}
\resizebox{\linewidth}{!}{
\begin{tabular}{lcccc}
\toprule
\multirow{2}{*}{\textbf{Method}} 
& \multicolumn{2}{c}{\textbf{Qwen2.5-7B}} 
& \multicolumn{2}{c}{\textcolor{black}{\textbf{Qwen3-4B(Thk)}}} \\ 
\cline{2-5}
& \textbf{WikiTQ} & \textbf{TabFact} & \textbf{WikiTQ} & \textbf{TabFact} \\ 
\midrule
No-Prep                  & 50.69          & 71.99          & \textcolor{black}{77.46}    & \textcolor{black}{92.89}    \\
Operation-R1             & \textbf{69.27} & \textbf{88.34} & \textcolor{black}{\textbf{81.63}}    & \textcolor{black}{\textbf{93.47}}    \\
$-$ ORPO \& VGR strategy & 63.49          & 85.67          & \textcolor{black}{73.87}    & \textcolor{black}{89.53}    \\
$-$ VGR strategy         & 68.67          & 86.35          & \textcolor{black}{80.57}    & \textcolor{black}{93.28}    \\
$-$ Operation Merge      & 68.16          & 85.67          & \textcolor{black}{76.84}    & \textcolor{black}{88.49}    \\
$-$ Adaptive Rollback    & 64.24          & 85.13          & \textcolor{black}{76.70}    & \textcolor{black}{87.35}     \\
\bottomrule
\end{tabular}
}
\vspace{-3.5mm}
\end{table}

\subsection{In-depth Analysis}

To gain deeper insights into how Operation-R1 achieves its performance improvements, we analyze the distribution and effectiveness of operator sequences across the WikiTQ and TabFact datasets.

\begin{table}[t]
\centering
\caption{Performance comparison on additional types of questions on TableBench and FeTaQA.}
\vspace{-1em}
\label{tab:additional_dataset}
\color{black}
\begin{tabular}{l l l c}
\toprule
\textbf{Dataset} & \textbf{Question Type} & \textbf{Method} & \textbf{Performance} \\
\hline
\multirow{8}{*}{TableBench}
& \multirow{4}{*}{\makecell{Fact checking}}
& No Prep & 57.29 \\
& & CoTable & 41.67 \\
& & AutoPrep & 65.62 \\
& & Operation-R1 & 77.08 \\
\cmidrule(lr){2-4}
& \multirow{4}{*}{\makecell{Numerical\\reasoning}}
& No Prep. & 11.59 \\
& & CoTable & 29.22 \\
& & AutoPrep & 50.76 \\
& & Operation-R1 & 44.87 \\
\hline
\multirow{4}{*}{FeTaQA}
& \multirow{4}{*}{\makecell{Free-form\\table QA}}
& No Prep. & 21.10 \\
& & CoTable & 19.89 \\
& & AutoPrep-CoT & 16.05 \\
& & Operation-R1 & 23.78 \\
\bottomrule
\end{tabular}
\vspace{-1.5em}
\end{table}

\noindent\underline{\textcolor{black}{\textit{Generalization to Unseen Datasets.}}}
\textcolor{black}{Table~\ref{tab:additional_dataset} reports cross-dataset results on TableBench and FeTaQA without any additional fine-tuning.
On TableBench, \textsf{Operation-R1} achieves the best fact-checking accuracy (77.08\%). On numerical reasoning, it surpasses No-Prep and CoTable but trails AutoPrep (44.87\% vs.\ 50.76\%), attributable to the absence of an \texttt{aggregate} operator. Introducing \texttt{aggregate} during inferenc (triggered in 52.64\% of NR instances) yields a +32.06\% gain on that subset (Table~\ref{tab:operator_ablation}), lifting overall NR accuracy to 61.96\% and surpassing all baselines.
On FeTaQA, \textsf{Operation-R1} achieves the highest BLEU score (23.78), while multi-step methods degrade below No-Prep. FeTaQA's multi-intent questions require joint reasoning over multiple sub-queries; sequential execution biases the pipeline toward one misdirection, whereas single-step generation captures global intent and avoids premature commitment.}

\noindent\underline{\textcolor{black}{\textit{Performance on Cell-Focused vs. Non-Cell-Focused Tasks.}}} 
\textcolor{black}{Although Operation-R1 is trained exclusively on cell-focused (CF) instances, Table~\ref{tab:cf_vs_ncf} shows that it yields consistent gains on non-cell-focused (NCF) tasks as well: on WikiTQ, the improvement over No-Prep on  (+23.02\%) even exceeds that on CF (+14.22\%), indicating that operator-based preprocessing effectively reduces reasoning burden regardless of whether answers are directly extractable as cell values. This trend is further confirmed on TabFact, which is inherently NCF, where Operation-R1 surpasses all baselines by substantial margins. These results confirm that the preprocessing policies learned from CF supervision transfer robustly to NCF settings.}

\begin{table}[ht]
\centering
\caption{\textcolor{black}{Performance comparison on Cell-Focused (CF) vs. Non-Cell-Focused (NCF) tasks.}}
\vspace{-1.5mm}
\label{tab:cf_vs_ncf}
\color{black}
\begin{tabular}{lllcc}
\toprule
\textbf{Dataset} & \textbf{Type} & \textbf{Prop.} & \textbf{Method} & \textbf{Accuracy (\%)} \\
\midrule
\multirow{8}{*}{WikiTQ}
& \multirow{4}{*}{CF} 
& \multirow{4}{*}{46.1\%}
& No-Prep & 63.29 \\
& & & CoTable & 56.72 \\
& & & AutoPrep & 60.77 \\
& & & Operation-R1 & \textbf{77.51} \\

& \multirow{4}{*}{NCF}
& \multirow{4}{*}{53.9\%}
& No-Prep & 39.92 \\
& & & CoTable & 48.28 \\
& & & AutoPrep & 54.80 \\
& & & Operation-R1 & \textbf{62.94} \\

\midrule
\multirow{4}{*}{TabFact}
& \multirow{4}{*}{NCF$^\dagger$}
& \multirow{4}{*}{100.0\%}
& No-Prep & 71.99 \\
& & & AutoPrep & 74.99 \\
& & & CoTable & 80.43 \\
& & & Operation-R1 & \textbf{88.44} \\
\bottomrule
\end{tabular}
\vspace{-1mm}
\end{table}

\begin{table}[t]
\centering
\caption{\textcolor{black}{LLM invocation comparison across Operation-R1 and baselines.}}
\vspace{-1em}
\label{tab:llm_calls}
\color{black}
\resizebox{\linewidth}{!}{
\begin{tabular}{l l c c c c}
\toprule
\textbf{Dataset} & \textbf{Method} & \textbf{Prepare} & \textbf{Extra} & \textbf{QA} & \textbf{Total Calls} \\
\midrule
\multirow{3}{*}{WikiTQ} 
& CoTable & 12537 & 197 & 4344 & 17078 \\
& AutoPrep & 9691 & 8109 & 5248 & 20965 \\
& Operation-R1 & 4344 & 134 & 4831 & 9309 \\
\midrule
\multirow{3}{*}{TabFact} 
& CoTable & 6258 & 44 & 2024 & 8326 \\
& AutoPrep & 3593 & 3780 & 2318 & 9691 \\
& Operation-R1 & 2024 & 56 & 2376 & 4456 \\
\bottomrule
\end{tabular}
}
\vspace{-2em}
\end{table}

\noindent\underline{\textcolor{black}{\textit{Analysis on LLM invocation.}}} 
\textcolor{black}{
Table~\ref{tab:llm_calls} breaks down the number of LLM invocations across three stages of TableQA: (1) \textit{Prepare}: generating data preparation pipelines; (2) \textit{Extra}: executing semantic operators such as \texttt{add\_column}; and (3) \textit{QA}: answering questions based on the curated table. \textsf{Operation-R1} reduces total LLM calls by over 46\% on both WikiTQ and TabFact compared to CoTable and AutoPrep, while maintaining superior accuracy. This reduction stems directly from our one-step pipeline generation: rather than invoking LLMs iteratively per operator as in loop-wise agentic methods, \textsf{Operation-R1} generates the entire pipeline in a single forward pass, keeping \textit{Prepare} calls equal to the number of test instances. The remaining overhead arises from a small number of parallel candidates used for Operation Merge and occasional calls for semantic operators, both of which are limited in frequency.
}

\noindent\underline{\textcolor{black}{\textit{Additional results.}}} 
\textcolor{black}{We report additional results including the prevalence of cell-focused instances across popular TQA benchmarks, a comparison of rollback strategies, a sensitivity analysis on the number of parallel inferences for Operation Merge, and scalability with table size in~\cite{li2026fullpaper} in the interest of space.}

\subsection{Generalization Ability}

\begin{table}[t]
\renewcommand\tabcolsep{1.45pt}
\centering
\caption{Accuracy comparison on subsets of datasets where the additional operator unseen in training is required for TableQA.}
\vspace{-2mm}
\label{tab:operator_ablation}
\begin{tabular}{lcccc}
\toprule
\multirow{2}{*}{Dataset} & \multirow{2}{*}{Usage (\%)} & \multicolumn{3}{c}{Accuracy (\%)} \\
\cmidrule(lr){3-5}
& & No Prep & ours & ours w/ additional \(op\) \\
\midrule
\multicolumn{5}{l}{\textcolor{black}{\textit{Subset requiring \texttt{aggregate} operator \quad (\(op = op_{\text{aggr}}\))}}} \\
\midrule
\textcolor{black}{TableBench-NR}  & \textcolor{black}{52.64} & \textcolor{black}{5.26}  & \textcolor{black}{37.80} & \textcolor{black}{\textbf{69.86 (+32.06)}} \\
\textcolor{black}{TableBench-FV}  & \textcolor{black}{13.54} & \textcolor{black}{46.16} & \textcolor{black}{61.54} & \textcolor{black}{\textbf{69.23 (+7.69)}}  \\
\midrule
\multicolumn{5}{l}{\textit{Subset requiring \texttt{clean\_column} operator \quad (\(op = op_{\text{clean}}\))}} \\
\midrule
WikiTQ  & 13.63 & 45.43 & 52.36 & \textbf{56.93 (+4.57)} \\
TabFact & 17.98 & 82.14 & 84.61 & \textbf{86.81 (+2.20)} \\
\bottomrule
\end{tabular}
\end{table}

\color{black}
\subsubsection{Generalization to extended operator space.} To explicitly test the scalability of our framework beyond the five operators seen during training, we add two operators only at inference time, targeting the plug-and-play extensibility of \textsf{Operation-R1}:
\color{black}

\textbf{\sql{Clean\_column.}}
The \texttt{clean\_column} operator performs simple value normalization such as fixing date formats or standardizing numeric units. The target column \(c\) and cleaning description \(d\) are treated as parameters of the operator, denoted
\(\text{clean\_column}_{c,d}\). The operator takes only a table \(T\) as input:
\begin{equation*}
    \text{clean\_column}_{c,d}(T)
\end{equation*}
and returns a table where column \(c\) has been normalized according to \(d\).
For example, \(\text{clean\_column}_{\text{Date},\,\text{`standardize date format'}}(T)\) normalizes date values in the Date column. 
This operator requires semantic understanding and is implemented by invoking an LLM.

\color{black}
\textbf{\sql{Aggregate}} supports group-by with numerical aggregations (e.g., \texttt{sum}, \texttt{mean}, \texttt{count}), parameterized by a grouping column \(c\), an aggregation target column \(c'\), and an aggregation function \(f\):
\begin{equation*}
    \text{aggregate}_{c,c',f}(T)
\end{equation*}
As shown in Table~\ref{tab:operator_ablation}, both operators yield consistent gains on the subsets where they are triggered. \texttt{aggregate}, required in 52.64\% of TableBench-NR instances, delivers a +32.06\% gain on that subset (lifting overall NR accuracy to 61.96\%) and a +7.69\% gain on TableBench-FV. \texttt{clean\_column}, triggered in 13--18\% of WikiTQ and TabFact instances, improves accuracy by +4.57\% and +2.20\% respectively. These results confirm that \textsf{Operation-R1} generalizes robustly to unseen operators—leveraging them appropriately without any retraining—and that cell-focused rewarding effectively incentivizes operator-based evidence localization and preservation.
\color{black}

\textcolor{black}{
\begin{table}[t]
\centering
\caption{\textcolor{black}{The evaluation results of \textsf{Operation-R1} trained on different datasets (Accuracy \%).}} 
\label{tab:train_data}
\color{black}
\begin{tabular}{l c c c}
\toprule
\textbf{Training set} & \textbf{Query Num.} & \textbf{WikiTQ Acc.} & \textbf{TabFact Acc.} \\
\midrule
WikiTQ  & 5403 & 69.66 & 88.44 \\
TabFact & 5000 & 69.77 & 88.49 \\
\bottomrule
\end{tabular}
\vspace{-4mm}
\end{table}
}

\vspace{-1em}

\subsubsection{\textcolor{black}{Generalization of the training regime to non-cell-focused tasks.}}
\textcolor{black}{
To show the generalization of our training regime, we adapt it to TabFact and evaluate the model on WikiTQ. To construct cell-focused supervision on TabFact, we prompt DeepSeek-V3.2 to annotate each claim-table pair with the \emph{terminal cells in the logical reasoning chain}—those cells whose values directly determine the verdict—as surrogate targets for the correctness reward. This preserves the per-operator verifiability required by ORPO while accommodating the non-extractive nature of the task. Following standard RL training practice~\cite{sun2026improvingdataefficiencyllm,chen2025unlockingpotentialdifficultyprior}, we apply difficulty-based filtering to construct a 5,000-example training set, excluding instances that are trivially solvable (by Qwen3-1.7B) or intractably complex (failing for Qwen3-30B-A3B).
}

\textcolor{black}{
As shown in Table~\ref{tab:train_data}, \textsf{Operation-R1} trained on this auto-annotated TabFact subset achieves performance nearly identical to its WikiTQ-trained counterpart on both benchmarks (69.77\% vs.\ 69.66\% on WikiTQ; 88.49\% vs.\ 88.44\% on TabFact). These results confirm that the cell-focused proxy is not a fundamental constraint but a \emph{generalizable training proxy}: for any table QA task, terminal reasoning cells can be identified via LLM annotation, enabling a plug-and-play extension to diverse table reasoning tasks without requiring explicit ground-truth cell annotations.
}

\section{RELATED WORK}

\subsection{Table Reasoning Paradigms}
\noindent\textbf{Text-based methods.} These methods serialize tables into natural language and apply in-context learning~\cite{chen-2023-large} or chain-of-thought reasoning~\cite{cot,ziqi2023tab}. Beyond generic prompting, extensions include decomposition over large tables~\cite{10.1145/3539618.3591708dater,guan2024mfort,yang2025triples} and structured linearization of complex headers~\cite{zhao2023large,ziqi2023tab}. These methods struggle with row-column structure encoding, incur high token overhead, and lack guarantees on numerical operations (e.g., sorting, aggregation).

\noindent\textbf{Program-based methods.} These methods prompt LLMs to produce executable SQL or Python code to guarantee numerical correctness. 
Early work such as Binder~\cite{Cheng2023Binding} adapts Program-of-Thought by binding LLM-generated symbolic programs to external verifiers. ReAct-style systems~\cite{Zhang2024ReAcTable} iteratively construct and execute partial programs with tool augmentation. Decomposition-based pipelines~\cite{10.1145/3539618.3591708dater,nguyen2024interpretable,wang2025tabsd} split complex queries into structured sub-programs before execution, while multi-step planning methods~\cite{Zhu2024AutoTQA} coordinate SQL generation across multiple reasoning stages and retrieval-augmented methods~\cite{jin2025talon} integrate external knowledge for program synthesis. 
Despite their strengths, these methods suffer from fragile program synthesis.

\noindent\textbf{Operator-centric methods.} These methods mediate between the two extremes by manipulating tables through composable operations, preserving structural reasoning while remaining simpler to generate than full programs. CoTable~\cite{wang2024chain-of-table-CoTable} treats iterative table transformation as the reasoning chain, while AutoPrep~\cite{fan2025autoprep} organizes a multi-agent (planner/programmer/executor) pipeline to produce a question-aware preprocessed table passed to a separate QA model.
\textcolor{black}{Compared to text2sql, such paradigms fit more in semi-structured and dirty tabular data and
provide better transparency and explainability.}
However, they require multiple LLM calls (up to 10) and super large models ($\ge$70B parameters), raising latency and cost concerns for real-time use.

\vspace{-1em}
\subsection{Learning Paradigms for Table Reasoning}

\noindent\textbf{Supervised End-to-End Training.} SFT-based methods optimize end-to-end table QA/fact verification with large curated or synthetic traces. Representative lines include TableGPT2~\cite{su2024tablegpt2largemultimodalmodel} and Table-R1-SFT~\cite{yang2025tabler1inferencetimescalingtable}, which achieve strong accuracy but require full-parameter training and tens of thousands of training signals (Table~\ref{tab:end2end}).

\noindent\textbf{RLVR for Table Reasoning.} RLVR methods~\cite{shao2024deepseekmath,mroueh2025reinforcement} optimize LLMs using automatically verifiable rewards. In the table domain, three Table-R1 threads exemplify different reward placements: (i) Yang et al. \cite{yang2025tabler1inferencetimescalingtable}  apply GRPO to short-form QA, fact verification, and free-form QA with task-specific rewards evaluating format adherence and answer correctness, enabling inference-time scaling without human annotations. (ii) Wu et al. \cite{wu2025tabler1teleai} and Lei et al. \cite{lei2025reasoningtable} introduced region-based reward mechanisms that compute the Intersection over Union (IoU) between predicted and ground-truth table regions (cells/columns/rows), using GRPO to optimize both evidence localization accuracy and final answer correctness. (iii) Jin et al. \cite{jin2025tabler1selfsupervisedreinforcementlearning} proposed a hybrid GRPO variant that dynamically switches between program-based and text-based reasoning using reasoning-type-specific rewards. 
However, current RLVR approaches either rely on sparse binary correctness signals~\cite{yang2025tabler1inferencetimescalingtable} or require manually annotated ground-truth regions and evaluate reasoning quality indirectly by checking whether models mention correct areas in natural language~\cite{wu2025tabler1teleai,lei2025reasoningtable}.
\vspace{-1mm}

\section{CONCLUSION}

In this work, we present \ours, enabling replacing multi-step data preparation pipeline assembly with one-step pipeline generation with a lightweight LLM (e.g., Qwen-1.7B/4B). To advance LLM's capability in pipeline generation, we propose ORPO, a LLM post-training algorithm with a novel self-supervised pipeline rewarding mechanism and a dynamic variance-aware group resampling strategy. Moreover, we introduce inference-time operation merge and adaptive rollback modules in \ours to enhance its robustness.
Extensive experiments on 4 TableQA datasets show that \ours improves TQA accuracy by 8.83\% and 4.44\% over multi-step preparation baselines with only 5K RL training examples, while reducing table size by 79\%, inference latency by up to 51\%, and 2.2$\times$ monetary cost, demonstrating the superiority of \ours in both TQA accuracy and efficiency.

\balance
\bibliographystyle{ACM-Reference-Format}
\bibliography{ref}

\clearpage
\appendix
\appendix
\section{Implementation Details}
\label{app:impl}

\subsection{Experiment Settings}
All experiments are conducted on a server equipped with 256 GB RAM, an Intel Xeon Silver 4314 CPU operating at 2.40 GHz, and an NVIDIA A40 GPU with 46 GB memory. The software environment includes PyTorch 2.4.0, Transformers 4.46.1, and the vLLM serving engine (version 0.6.2). To ensure reproducibility, we fix the random seeds across all frameworks. Due to GPU memory limitations, all vLLM-based experiments use half-precision (FP16) inference. \vspace{1.5mm}

\noindent\textbf{Training protocol.} Our training for the operation generator is conducted only on the Cell-Focused subset of WikiTQ, \textcolor{black}{whose question-table pairs are disjoint from the test set by dataset construction}, while TabFact is reserved purely as an out-of-domain evaluation benchmark to assess the transferability of the learned data-preparation policies.

\subsection{Training Configuration}
We develop Operation-R1 models using Qwen3 series backbones: Operation-R1-1.7B on Qwen3-1.7B and Operation-R1-4B on Qwen3-4B. 
For ORPO training described in Section~\ref{sec:setup}, we apply Cell-Focused QA filtering to WikiTQ and keep only instances shorter than 2,800 tokens\textcolor{black}{---imposed by our GPU memory constraints (A40, 46G)--}yielding 5,403 examples from the original training set.
For fair comparison, the GRPO and DAPO baselines in Table~\ref{tab:end2end} are trained on an 11K subset from the same split, filtered only using the same 2,800-token length constraint. We adopt a 19:1 train-validation split and utilize the ms-swift framework with the following hyperparameters: maximum sequence length of 5{,}632 tokens, rollout parameter of 12, total batch size of 144, and learning rate of \(7 \times 10^{-7}\). For the length-aware reward, we set \(L_{\text{max}} = 2{,}560\) and \(L_{\text{cache}} = 512\). In the VGR strategy, we use a variance threshold of 0.1 and a quality threshold of 0.5. The LoRA rank is set to 32, and we enable Liger kernel optimization along with Flash Attention for efficient training. DAPO-related parameters follow the configuration specified in the literature ~\cite{yu2025dapoopensourcellmreinforcement}. 
\textcolor{black}{\textsf{Operation-R1}-1.7B training takes 54 GPU-hours on 2 NVIDIA A40 GPUs; \textsf{Operation-R1}-4B takes 114 GPU-hours on 4 NVIDIA A40 GPUs.}

\section{More Experiments}

\subsection{Analysis on Cell-Focused vs.\ Non-Cell-Focused Tasks}
\label{app:cf_ncf}

\begin{table}[t]
\centering
\caption{Quantitative breakdown of popular TQA benchmarks by cell-focused (CF) instances.} \vspace{-3mm}
\label{tab:cf_ncf_benchmarks}
\begin{tabular}{l r c}
\toprule
\textbf{Benchmark} & \textbf{Task Type} & \textbf{CF (\%)} \\
\midrule
WikiTQ  & Short-form QA & 46.1 \\
HybridQA & Hybrid QA & 40.4 \\
SQA      & Sequential QA  & 100 \\
KET-QA   & KB-Augmented QA & 44.0 \\
WikiSQL  & NL2SQL & 71.1 \\
\bottomrule
\end{tabular}
\vspace{-4mm}
\end{table}

\noindent\textbf{Prevalence of Cell-Focused Tasks.}
Table~\ref{tab:cf_ncf_benchmarks} reports the fraction of cell-focused (CF) instances across popular TQA benchmarks. CF instances constitute a non-trivial proportion in every benchmark surveyed, ranging from 40.4\% in HybridQA to 100\% in SQA. This indicates that cell-focused supervision signals are readily obtainable from existing datasets without requiring special annotation efforts, making the CF assumption a practical and broadly applicable training proxy rather than a restrictive constraint.

 \begin{figure}[t]
  \centering
  \includegraphics[width=\linewidth]{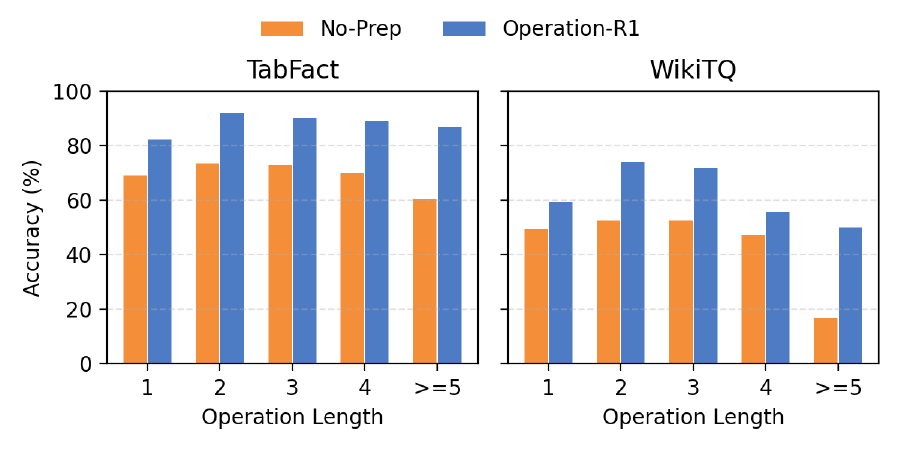}
   \vspace{-8mm}
  \caption{Impact of Operator Sequence Length on WikiTQ and TabFact.}
  \label{fig:length}
  \vspace{-4mm}
\end{figure}

\begin{figure}[t]
  \centering
  \includegraphics[width=\linewidth]{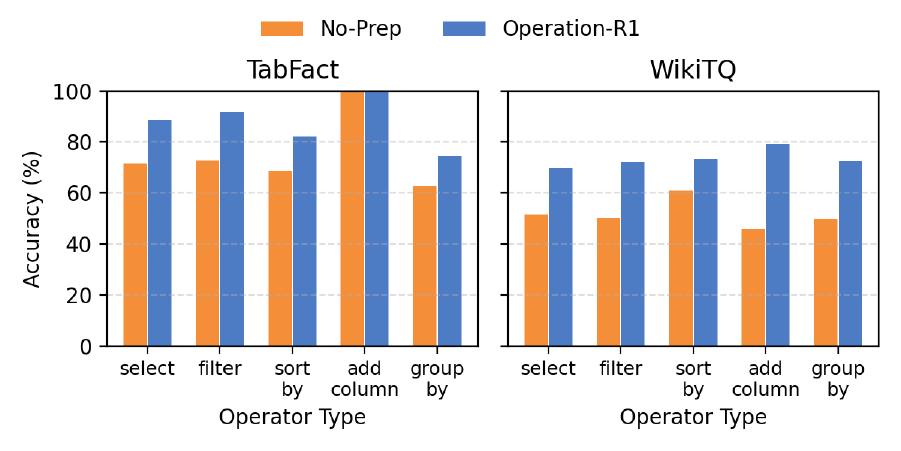}
  \vspace{-8mm}
  \caption{Accuracy by Operator Type on WikiTQ and TabFact.}
  \label{fig:ops}
  \vspace{-3mm}
\end{figure}

\noindent\textbf{Impact of Operator Sequence Length.} Fig.~\ref{fig:length} shows that as operator sequence length increases, \textsf{Operation-R1}'s advantage over the baseline widens. Longer sequences correspond to more complex questions requiring deeper table restructuring; the growing performance gap reflects \textsf{Operation-R1}'s ability to precisely transform raw tables into question-oriented representations. This trend also validates ORPO training: rather than plateauing on harder instances, improvements amplify with complexity—contrasting sharply with direct natural language reasoning, which degrades disproportionately as questions become more demanding.

\noindent\textbf{Operator Type Distribution and Effectiveness.} Fig.~\ref{fig:ops} shows accuracy gains broken down by operator type, where each question is bucketed by the operators in its predicted pipeline. \texttt{filter} and \texttt{select\_column} yield the largest gains (22.1\%/19.0\% and 18.1\%/17.1\%), reflecting \textsf{Operation-R1}'s effectiveness at eliminating irrelevant rows and columns—the classic "needle in a haystack" challenge. \texttt{sort\_by} shows moderate but consistent improvements (12.4\%/13.6\%) on comparison-intensive queries, while \texttt{group\_by} exhibits higher variance (22.8\%/11.9\%) owing to the more specialized demands of aggregation tasks. Cross-dataset patterns further confirm adaptability: TabFact benefits most from \texttt{filter} for evidence localization, while WikiTQ gains more uniformly across operator types, consistent with its diverse reasoning demands.

\subsection{Analysis on Rollback Strategy}
\label{app:invocation}

\begin{table}[t]
\centering
\caption{Rollback trigger statistics of Operation-R1 during inference. 
Step1–Step3 denote the pipeline length (number of operators) when rollback is triggered.}
\label{tab:rollback_stats}
\begin{tabular}{l c c c c c}
\toprule
\textbf{Dataset} & \textbf{No trigger} & \textbf{Full} & \textbf{Step3} & \textbf{Step2} & \textbf{Step1} \\
\midrule
WikiTQ  & 4001 & 343 & 7 & 92 & 45 \\
TabFact & 1715 & 309 & 2 & 54 & 37 \\
\bottomrule
\end{tabular}
\end{table}

\noindent\textbf{Rollback Trigger Statistics.}
Table~\ref{tab:rollback_stats} reports how frequently the adaptive rollback mechanism is triggered during inference. Rollback occurs only in a minority of cases: on WikiTQ, 92.1\% of instances require no rollback, and only 1\% of instances reach the final fallback state; on TabFact, 84.7\% require no rollback and only 1.8\% reach the final state. In cases where rollback is triggered, most require only a single rollback step, confirming that the binary rollback strategy efficiently locates the nearest recoverable state without excessive overhead. These results indicate that adaptive rollback functions as a lightweight guardrail—invoked sparingly to handle edge cases of over-processing—rather than a component the system routinely depends on.

\begin{table}[t]
\centering
\caption{Comparison of rollback strategies (Accuracy \%).}
\vspace{-1mm}
\label{tab:rollback_strategy}
\resizebox{\linewidth}{!}{
\begin{tabular}{l l c c}
\toprule
\textbf{Dataset} & \textbf{Strategy} & \textbf{Accuracy} & \textbf{Extra LLM Calls} \\
\midrule
\multirow{3}{*}{WikiTQ}
& Full-First-Origin & 69.27 & 431 \\
& Binary rollback (ours) & \textbf{69.66} & 487 \\
& Step-by-step rollback & 68.95 & 941 \\
\midrule
\multirow{3}{*}{TabFact}
& Full-First-Origin & 88.34 & 338 \\
& Binary rollback (ours) & \textbf{88.44} & 402 \\
& Step-by-step rollback & 87.35 & 483 \\
\bottomrule
\end{tabular}
}
\end{table}

\noindent\textbf{Rollback Strategy Explore.}
We compare three rollback strategies in Table~\ref{tab:rollback_strategy}. \textsf{Full-first-origin} always rolls back to the state after the first operator $\mathit{op}_1$, or to the original table if necessary, incurring minimal overhead. \textsf{Step-by-step} rollback reverts one operator at a time until a valid answer is obtained, providing fine-grained recovery at the cost of substantially more LLM calls (941 on WikiTQ). 
Binary rollback achieves the highest accuracy on both datasets (69.66\% and 88.44\%) with a moderate number of extra LLM calls (487 and 402), avoiding the over-aggressive truncation of \textsf{full-first-origin} and the excessive overhead of step-by-step recovery. We therefore adopt \textsf{binary rollback} as the default strategy.

\subsection{Analysis on Operation Merge}

\begin{figure}[t]
  \centering
  \includegraphics[width=\linewidth]{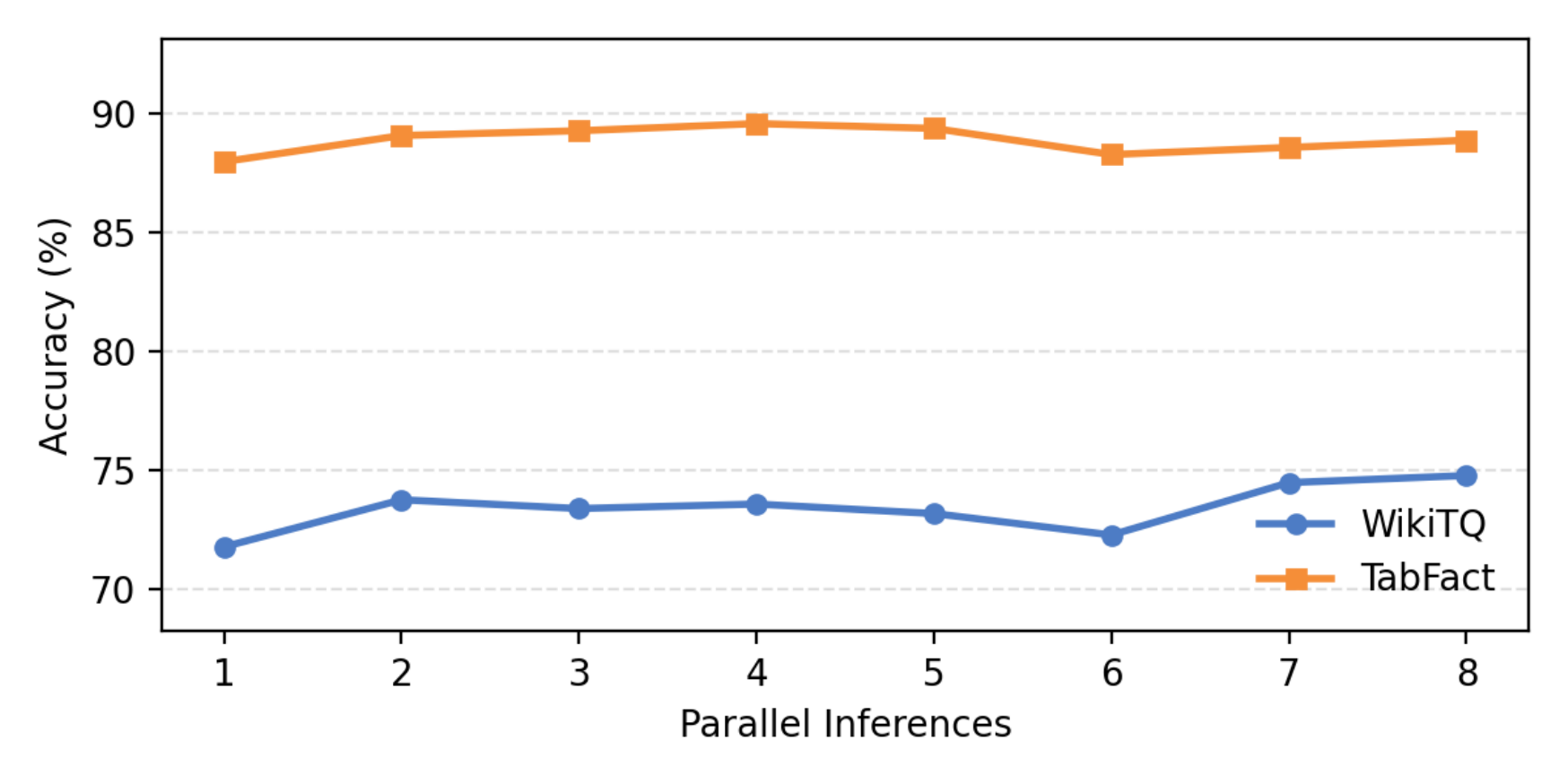}
   \vspace{-8mm}
  \caption{Impact of parallel inference count on Operation Merge.}
  \label{fig:merge_n}
  \vspace{-4mm}
\end{figure}

Fig.~\ref{fig:merge_n} illustrates how the number of parallel inferences ($N$) affects Operation-R1's accuracy on WikiTQ and TabFact. Both datasets exhibit a steady performance increase as $N$ scales from 1 to 4, with accuracy rising from 71.80\% to 73.60\% and 88.00\% to 89.60\%, respectively, confirming that consensus-based merging over a growing candidate pool effectively filters spurious operations. Beyond $N=4$, however, accuracy fluctuates without consistent gains, indicating diminishing returns from additional inferences. We therefore adopt $N=4$ as the default, striking a favorable balance between accuracy and inference overhead.

\subsection{Scalability with Table Size}
\label{app:table_size}

\begin{figure}[t]
  \centering
  \includegraphics[width=\linewidth]{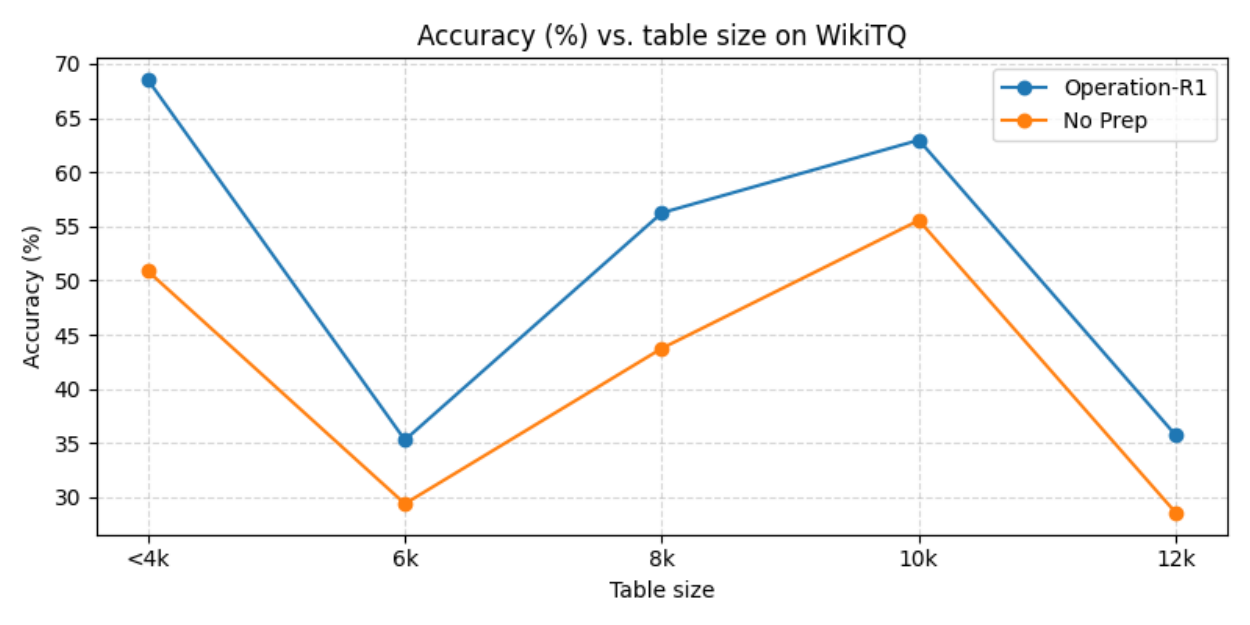}
   \vspace{-8mm}
  \caption{Accuracy comparison across different table lengths (measured by serialized token count).}
  \label{fig:table_length}
  \vspace{-2mm}
\end{figure}

Figure~\ref{fig:table_length} reports accuracy as a function of table size, measured by serialized token count. Both \textsf{Operation-R1} and No-Prep degrade as table size grows, reflecting the increasing difficulty of reasoning over longer contexts. However, \textsf{Operation-R1} consistently outperforms No-Prep across all scales, and the advantage is often amplified at larger sizes—for example, the gain reaches $+12.50\%$ ($43.75\% \to 56.25\%$) at 8K tokens and $+17.65\%$ at 6K tokens. This trend confirms that operator-based preprocessing becomes increasingly beneficial as table size grows, as explicit compression directly alleviates the ``needle-in-a-haystack'' problem that deteriorates raw LLM reasoning on large tables.

\subsection{Analysis on Thinking Mode}
\label{app:thinking}

Table~\ref{tab:qwen_thinking} compares CoTable under Qwen3-4B thinking and non-thinking modes. Although thinking mode improves accuracy by +4.77\% on WikiTQ, it increases inference time by $8\times$ (from 8,008s to 65,837s), yielding a poor cost-benefit trade-off for multi-step operator-centric frameworks where repeated LLM invocations already constitute the primary latency bottleneck. Extending thinking mode to all baselines would furthermore require non-trivial adaptation of their prompting pipelines. For these reasons, Table~\ref{tab:main_results} reports thinking-mode results only for No-Prep and CoTable.

\begin{table}[t]
\centering
\caption{Performance comparison using Qwen3-4B under thinking and non-thinking modes.}
\label{tab:qwen_thinking}
\resizebox{\linewidth}{!}{
\begin{tabular}{l l cc cc}
\toprule
\multirow{2}{*}{\textbf{Method}} & \multirow{2}{*}{\textbf{Model}} 
& \multicolumn{2}{c}{\textbf{WikiTQ}} 
& \multicolumn{2}{c}{\textbf{TabFact}} \\
\cmidrule(lr){3-4} \cmidrule(lr){5-6}
& & \textbf{Acc. (\%)} & \textbf{Time (s)} & \textbf{Acc. (\%)} & \textbf{Time (s)} \\
\midrule
\multirow{2}{*}{CoTable}
& Qwen3-4B & 57.50 & 8008 & 86.46 & 4743 \\
& Qwen3-4B (Thk) & 62.27 & 65837 & 89.62 & 25392 \\
\bottomrule
\end{tabular}
}
\end{table}

\end{document}